\documentclass[aps,pra, twocolumn,nofootinbib, showpacs]{revtex4-1} 
\usepackage[unicode=true,pdfusetitle, bookmarks=true,bookmarksnumbered=false,bookmarksopen=false, breaklinks=false,pdfborder={0 0 0},backref=false,colorlinks=false] {hyperref}
\hypersetup{ colorlinks,linkcolor=myurlcolor,citecolor=myurlcolor,urlcolor=myurlcolor}
\usepackage{braket,colortbl,amsthm,amsmath,cleveref,amssymb,txfonts}\definecolor{myurlcolor}{rgb}{0,0,0.7}
\usepackage{graphics,graphicx}
\usepackage{color}
\usepackage[usenames,dvipsnames]{xcolor}
\usepackage{amssymb}
\usepackage{amsthm}
\usepackage{amsfonts}
\usepackage{float}
\usepackage{graphicx}
\usepackage[format = plain,labelfont = bf,up, textfont = normal , up, justification =raggedright, singlelinecheck =false]{caption}
\usepackage{subcaption}
\usepackage{tabularx}

\usepackage{graphicx}
\usepackage[utf8x]{inputenc}
\usepackage{color,soul}
\usepackage{amsmath}
\usepackage{braket}
\usepackage{latexsym}
\usepackage{bm}
\usepackage{graphics,epstopdf}
\usepackage{enumitem}
\usepackage{fmtcount}
\usepackage{booktabs}
\usepackage{csquotes}
\usepackage{epsfig}
\theoremstyle{plain}

\def\bea{\begin{eqnarray}}
\def\eea{\end{eqnarray}}
\def\ba{\begin{array}}
\def\ea{\end{array}}

\def\beq{\begin{equation}}
\def\eeq{\end{equation}}

\begin{document}


\title{Response to glassy disorder in coin on spread of quantum walker}

\author{Priya Ghosh, Kornikar Sen, Ujjwal Sen}

\affiliation{Harish-Chandra Research Institute, A CI of Homi Bhabha National Institute, Chhatnag Road, Jhunsi, Allahabad 211 019, India}
\begin{abstract}

We analyze the response to incorporation of glassy disorder in the coin operation of a discrete-time quantum walk in one dimension. We find that the ballistic spread of the disorder-free quantum walker is inhibited by the insertion of disorder, for all the  disorder distributions that we have chosen for our investigation, but remains faster than the dispersive spread of the classical random walker. Beyond this generic feature, there are significant differences between the responses to the different types of disorder. In particular, the falloff from ballistic spread can be slow (Gaussian)  or fast (parabolic) for different  disorders, when the strength of the disorder is still weak. 
The cases of slow response always pick up speed after a point of inflection at a mid-level disorder strength. The disorder distributions chosen for the study are Haar-uniform, spherical normal, circular, and two types of spherical Cauchy-Lorentz.
\end{abstract}
\maketitle

\section{Introduction}
\label{sec1}


In a classical random walk~(CRW) on a line, the walker chooses to move left or right, depending on the 
result of a coin toss, and a pre-decided correlation between the two degrees of freedom.
%
After a large number of steps, the position distribution of the walker is Gaussian and the corresponding spread, as quantified by the standard deviation, is proportional to the square root of the number of steps.  
A quantum version~\cite{ref1} of the classical random walk 
%
has a standard deviation of the position distribution of the walker that is linear in the number of steps~\cite{ref2, ref3}.
Quantum walks~(QWs) have since received a lot of attention~\cite{ref4,ref140,ref5,ref6,ref7,ref8,ref9,add6},
%
%
%
and 
have been applied to a plethora of scenarios, such as quantum algorithms~\cite{ref9,ref10,ref11,
ref12,ref13}, quantum memory~\cite{ref14,ref15}, and so on~\cite{ref2,ref4,ref16,ref17}.
They have also been used to model 
quantum 
state transfer~\cite{ref18,ref19,ref20,ref21}. QWs can  - in the main - be classified into two categories, viz. discrete-~\cite{ref1,ref22} and continuous-time~\cite{ref5,ref16,ref23}
ones.
%
%
%
Quantum walks have been experimentally realized in nuclear magnetic resonance systems, photons in waveguides, trapped atoms, synthetic gauge fields in a three-dimensional lattice, Fibonacci fibers, superconducting qubits, etc.~\cite{add1,ref32,ref33,ref34,ref35,ref36,ref37}. 
 
 Quantum walks have been discussed on different types of graphs such as Cayley graphs, percolation graphs, glued trees graphs, cyclic graphs, etc. for discrete~\cite{add2,ref38,ref39,ref41} as well as continuous~\cite{add3,add4,add5,ref44} quantum walks. Topological behavior (topological invariants and topological phases) of quantum walks has also been illustrated, both theoretically~\cite{add6,add7,add8,add9} and experimentally~\cite{ref47,ref48,ref49,ref50,add10}. 
 Quantum walks in presence of decoherence and jumps were also studied
 \cite{ref51,ref52,ref53,ref54}.
 It has been used as a powerful tool in simulations of 
 physical processes such as 
 photosynthesis, breakdown of electric-field-driven systems, and so on~\cite{ref55,ref56,ref58}. 

Imperfections are ubiquitous in  physical systems, and it is important to consider their effects in  physical phenomena and devices. While imperfections in the form of disorder or noise are typically expected to reduce the visibility in the order parameter of a phenomenon and the efficiency of a device, it has surprisingly been also found to increase visibilities and efficiencies in certain cases~\cite{ref76,ref77,ref78,ref79,ref80,ref81,ref82}.  
It is therefore natural to consider the effects of disorder insertion in quantum walks~\cite{ref46,ref59,ref60,ref61,ref62,ref63,ref64,ref65,ref66,ref83,ref84,ref85,ref86,ref87,ref88,ref89,ref90,ref91,ref92,ref93,ref94,ref95,ref96,ref97,ref98,ref99,ref100,ref101,ref102,ref103,ref104,ref105,ref106,ref107,ref108}.
%
It has been seen that the spread of a quantum walker can get reduced in the presence of noise. Imperfections have been considered in the quantum coin~\cite{ref92,ref94,ref95,ref96,ref100,ref101,ref102,ref103,ref104,ref105,ref106,ref107,ref108} as well as in the step length of the walker~\cite{ref63,ref99}. At least three types of disorder have been identified and imposed on the quantum coin - static (position dependent coin)~\cite{ref92,ref100,ref101,ref102,ref103}, dynamic/temporal (time dependent coin)~\cite{ref95,ref96,ref104,ref105,ref106,ref107,ref108}, fluctuating disorder (coin as a function of both time and position)~\cite{ref94}. 
%
%

In this work, we focus on one-dimensional (1D) discrete-time quantum  walks (DQWs). We explore the spreading of the quantum walker in presence of ``glassy'' or ``quenched'' disorder in the coin operator. 
We find that the disorder-averaged spread of the position distribution of the quantum walker is slower than the case when there is no disorder. We quantify the spread by using the standard deviation of the position distribution of the walker. While the inhibition is common to all the distributions considered, there are clear differences in the behaviors of the scaling exponents of the spread for different distributions, and sometimes the difference can be even qualitative. In particular, the fall-off, from the ballistic spread in the disorder-free case, for increasing disorder that is 
still weak
can be slow (Gaussian) or fast (parabolic).
The slow response cases always have a point of inflection for a higher strength of disorder, after which the response picks up speed.
In most of the cases, for strong disorder, the spread is near-dispersive, 
but universally remains super-dispersive.
The classical random walker has a dispersive spread.
The disorder distributions considered are Haar-uniform with a finite cut-off, spherical normal (von Mises - Fisher), circular, and two spherical Cauchy-Lorentz distributions. 

The structure of the remaining parts of the paper is 
as follows. In Sec.~\ref{sec2}, we briefly  recapitulate some general aspects of 1D DQWs. Sec.~\ref{sec3} consists of a concise discussion on glassy disorders in general, and the model of the coin's disorder that we use.  We then briefly discuss, in Sec.~\ref{sec5}, the different types of probability distributions on a sphere that we use for prototyping the disorder distribution in the coin operator. 
The momentum representation is considered in 
Sec.~\ref{sec4}, the analytical considerations in which help in the numerical analysis beyond. 
The scaling analysis is presented 
in the remaining portion of Sec.~\ref{sec6}. A conclusion is given in Sec.~\ref{sec7}.

\section{About quantum walk}
\label{sec2}
The 1D discrete-time CRW is a stochastic process which represents the path of a ``drunk'' walker. It consists of taking a step to the left or right of the current position, with the direction being 
decided by another degree of freedom, a ``classical coin'', and this process tossing the coin and moving a step according to the result of the toss is repeated many times. 
Brownian motion of gas or liquid molecules has been modelled by 
CRWs \cite{ref109,ref110}.

In the quantum analogue of discrete-time CRWs, 
both the walker and the coin are described as quantum systems, and they are typically quantum coherent~\cite{ref111,ref112,ref113,ref114,ref115,ref116,ref117}
and entangled~\cite{ref118,ref119,ref120,ref121}
at almost all instants. 
We denote the Hilbert space of the coin by $\mathcal{H}_c$, with the elements of the computational basis,
$\{\ket{0},\ket{1}\}$, representing ``head'' and ``tail'' in a ``coin toss''. As we will see, the coin toss in a quantum walk is ``coherent'' and not a measurement in the computational basis. 
The Hilbert space of the walker, on the other hand, is denoted by $\mathcal{H}_w$, with the set of  positions of the walker in different steps, 
$\{\ket{i}:i\in 
\mathcal{Z} \}$, spanning it, where $\mathcal{Z}$ is the set of integers. 
Although the dimension of $\mathcal{H}_w$ is 
\(\aleph_0\),
after $t$ steps taken by the walker, the state of $\mathcal{H}_w$, written in the position basis will have nonzero coefficients only for the basis elements \(\ket{i}\), with \(i\in [-t,t]\cap \mathcal{Z}\), if the initial position is at \(|0\rangle\). The Hilbert space of the total system consisting of the walker and the coin is $\mathcal{H}_{tot}=\mathcal{H}_c\otimes\mathcal{H}_w$.


In DQWs, the walker's direction for each movement will be decided coherently by the  state of the coin at that step.
We assume that if the coin is in the state, $\ket{0}$, ($\ket{1}$), in any particular step, then the walker will move one step towards the right (left). The tossing of the coin is represented using the coin operator, $C$, which is a unitary gate acting on the Hilbert space $\mathcal{H}_c$, whereas the evolution of the walker is achieved by 
the conditional shift operator, $\tilde{S}$, on $\mathcal{H}_{tot}$, given by \begin{equation}
\tilde{S} = \sum_{i=-t}^{t}  \left(\ket{0} \bra{0} \otimes \ket{i+1} \bra{i} + \ket{1} \bra{1} \otimes \ket{i-1} \bra{i}\right).
\label{ma-tui-emon-kyano-holi}
\end{equation}
Hence if the initial state of the composite system is $\ket{\psi_{tot}}$, then the  state after one iteration, i.e., after taking one step is given by \(\ket{\psi_{tot}^{\prime}} = \tilde{S}(C\otimes I)\ket{\psi_{tot}}\).
Note that both CRWs and DQWs  have the parity property, in that  after an even (odd) number of steps, the probability of finding the walker at odd (even) displacements is zero. 
But in the quantum case, the variance of the position distribution of the walker varies quadratically with time, i.e., with the number of steps.


In DQWs, one typically uses the Hadamard gate, 
%
%
$$ H = \frac{1}{\sqrt{2}}\left (\begin{array}{cc}
1 & 1  \\
1 & -1 \\
\end{array} \right)$$
for the coin tossing operator, \(C\), and in such case, the walk may be referred to as the 
``Hadamard walk".
The position distribution can be asymmetric, with symmetric ones obtained by using the initial state of the coin as 
%
%
\[\ket{\psi_c^{in}} = 
\left(\ket{0} + i \ket{1} \right)/\sqrt{2},\]
or
by using the operator,
$$ Y = \frac{1}{\sqrt{2}}\left (\begin{array}{cc}
1 & i  \\
i & 1 \\
\end{array} \right),$$
as the coin operation, \(C\).
In this work, we consider the initial states of the quantum coin and the walker as \(\ket{\psi_c^{in}}\) and \(|0\rangle\) respectively. 
We set 
$$ \ket{\psi_{tot}^{in}} = \ket{\psi_c^{in}} \otimes \ket{0}.$$ 
The coin operator will be chosen as one that is Hadamard ``on average, but with a spread''. The meaning of this statement will be made precise below.

\section{Disorder in coin operation}
\label{sec3}
We wish to examine the effect of incorporation of a ``glassy'' disorder on the spreading of the quantum walker. The disorder is called ``glassy'', as the typical  equilibration time of the disorder is assumed to be several orders of magnitude higher than the typical observation times that we are interested in. This type of disorder is also called ``quenched'' disorder in the literature.

We introduce the disorder on the quantum coin in such a way that instead of transforming the state $\ket{0}$ ($\ket{1}$) to 
$(\ket{0}+\ket{1})/\sqrt{2}$ $\left((\ket{0}-\ket{1})/\sqrt{2}\right)$, the coin operator projects $\ket{0}$ ($\ket{1}$) to the state, $\ket{\xi(\theta,\phi)}$ $\left(\ket{\xi^\perp(\theta,\phi)}\right)$, on the surface of the Bloch sphere. The orthonormal states $\ket{\xi(\theta,\phi)}$ and $\ket{\xi^\perp(\theta,\phi)}$ can be described using the spherical polar angles, $\theta,~\phi$. 
The disorder chooses the values of $(\theta,\phi)$ by following certain probability distributions. We will discuss later about the choice of the probability distribution. 
We name such a coin operator as a \textit{biased} Hadamard gate. Such a gate is of course not strictly ``biased'', unless \(\phi \ne 0\), which, however, happens almost all the time. 

As stated before, we wish to consider the effect of glassy disorder on the QWs. More precisely, we wish to analyze the response to disorder insertion in the spread of the walker. At any particular step, the coin operator is chosen by randomly choosing the coin-operator parameters, \(\theta\), \(\phi\), from a pre-decided probability distribution. The choice at any particular step is made independently of the choice in any other step. We are interested in the spread of walker, as quantified by the standard deviation of the position distribution. For a given set of configurations of the disorder in a particular run, we calculate the spread, and then average over the disorder. This mode of averaging of a physical quantity \emph{after} its evaluation, has been called ``quenched'' averaging in the literature.   
%
%
%
Let the state of the composite system, consisting of the walker and the coin, after $t$ steps be $\ket{\psi_{tot}^t}$. Then the state of the walker after \(t\) steps is $$\rho_t=\text{Tr}_{c}(\ket{\psi_{tot}^t}\bra{\psi_{tot}^t}),$$ where $\text{Tr}_c$ denotes the partial trace taken over the coin's Hilbert space, $\mathcal{H}_c$. Then the variance of the position of the walker is given by
\begin{equation*}
    \sigma^2(\rho_t)=\sum_i \bra{i}\rho_t^2\ket{i}-\left(\sum_i\bra{i}\rho_t\ket{i}\right)^2.
\end{equation*}
We denote the disorder averaged standard deviation of the walker's position 
after $t$ steps as $\langle \sigma(t) \rangle$.

\section{Classic probability distributions on sphere}
\label{sec5}

For the situation that we are considering, when the coin is without any disorder, the coin operator is the Hadamard gate, which takes the \(\sigma_z\) eigenbasis onto the 
\(\sigma_x\) eigenbasis. 
In the disordered case, the target basis is \(\{|\xi(\theta,\phi)\rangle, |\xi^\perp(\theta,\phi)\rangle\}\), with a certain distribution for the spherical polar coordinates, \(\theta,~\phi\). 
In a realistic case, we expect this target basis to be close to the \(\sigma_x\) eigenbasis and distributed symmetrically around it.   We are therefore looking for probability distributions that are symmetrically distributed around \((\theta,\phi) = (\pi/2,0)\) on the unit sphere, to define the state \(|\xi(\theta,\phi)\rangle\). This automatically fixes the state \(|\xi^\perp(\theta,\phi)\rangle\), except for a phase, which is arbitrarily chosen to be such that the biased Hadamard gate is \(H(\theta, \phi)\), to be explicitly defined in Eq.~(\ref{shapla}).




We are mainly interested in the disorder averaged dispersion,  $\langle\sigma(t)\rangle$, of the disordered quantum walk. We want to study how  $\langle\sigma(t)\rangle$ depends on the number of steps, $t$. In particular, we want to estimate the ``scaling'', with the number of steps, of average dispersion of the disordered quantum walk, and its behavior with respect to the ``strength'' of the disorder. The meaning of the strength of a disorder will depend on the disorder distribution chosen, as we will see below. The scaling of the dispersion is defined as \(\alpha\), where
\begin{equation}
\label{ekTa-gaan-likho}
\langle\sigma(t)\rangle \propto t^{\alpha}.
\end{equation}
It is known that \(\alpha = 1\) in case of the ordered quantum walk, whereas in case of the classical random walk, it is 0.5. To find the value of $\alpha$ in a disordered quantum walk, we numerically evaluate $\braket{\sigma(t)}$ for different iterations, $t$, from 8 to 24, with intervals of 2, for each type of probability distribution of the disorder that we consider. These evaluations, obtained through log-log scaling plots, are correct up to two significant figures. 






From now on, we will use the following notation. When a point in the 3D space is expressed using the Cartesian coordinate system, we will use ``$c$" in the suffix of the vector, 
and when the coordinate system is spherical polar, we will use ``$s$" in the suffix. For example, $(r,\theta,\phi)_s$ is the spherical polar representation of a point, whereas the corresponding Cartesian representation is $(x,y,z)_c$
where $x=r\sin{\theta}\cos{\phi}$, $y=r\sin{\theta}\sin{\phi}$ and $z=r\cos{\theta}$.

To quantify the strength of disorder, we introduce the quantity, \(\sigma_{dis}\), that is essentially the standard deviation the disorder distribution, except that we remember that the distribution is scattered over a curved - and not flat - surface (the Bloch sphere):
\begin{equation}
    \sigma_{dis}=\left(\int dist(P,P_0) p(\theta_P,\phi_P)d\theta_P. d\phi_P\right)^{1/2}\label{eq5}
\end{equation} Here $P=(\sin\theta_P\cos\phi_P,\sin\theta_P\sin\phi_P,\cos\theta_P)_c$ is an arbitrary point on the surface of the sphere obtained from the disorder distribution, where $\theta_P$ and $\phi_P$ are zenith and azimuth angles respectively. And $P_0$ is the point on the Bloch sphere in the direction of the mean of the disorder distribution. 
The function, $dist(P,P_0)$, denotes the length of the shortest curved line, joining the points $P$ and $P_0$, drawn over the surface of the  sphere, and $p(\theta_P,\phi_P)$ is the probability density function. The integration is to be evaluated over the entire range of the disorder. The disorders considered here are taken to be distributed around the point $(1,0,0)_c$ on the Bloch sphere. Thus we take $P_0=(1,0,0)_c$ for all disorder distributions. The value of $\sigma_{dis}$ is found analytically for uniform and circular distributions, while for others, it is found numerically. We discuss now about the various distributions for disorders that we will consider. The suffix ``dis'' will be replaced by the disorder distribution utilized.

\subsection{Uniform distribution}
\label{subsec1}
\begin{figure}[h!]
\includegraphics[scale=0.5]{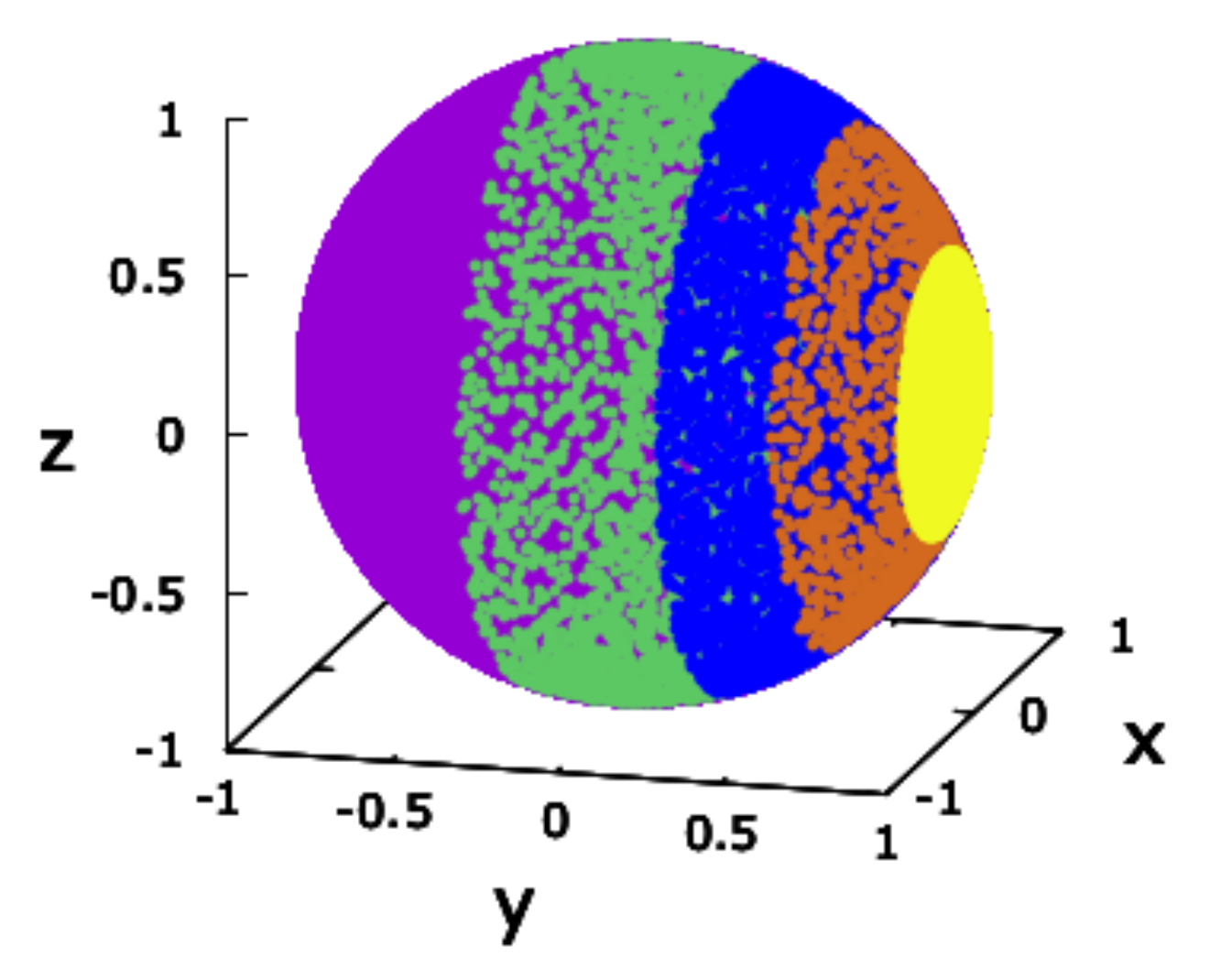}
\caption{Uniform distribution on a sphere. We exhibit here uniform distributions of points for different ranges, on a purple unit sphere, with the mean direction of the distribution being along $(1,0,0)_c$.
The differently colored dots, \textit{viz.} yellow, orange, blue, and green are chosen from uniform distributions with disorder strength (\(\sigma_U\)) 0.318, 0.648, 0.874, and 1.25 respectively. The corresponding values of $r_U$ are 0.1 (yellow), 0.4 (orange), 0.7 (blue), and 1.3 (green) respectively. [On grayscale, the different distributions for the different disorder strengths can be read from the figure by looking at their different ranges covered, and this is also true for  Figs.~\ref{fig2},~\ref{fig3},~\ref{fig4},~\ref{fig5}.]
There are $10^4$ 
points for each range. The uniform generations utilize the the method 
of Haar-uniform generation on a sphere. All quantities used are dimensionless.} 
\label{fig1} 
\end{figure}

We begin with a distribution for which 
the 
points, $(1,\theta,\phi)_s$, are uniformly distributed on the 
surface of a unit sphere (with center at the origin). As we have mentioned previously, we want a  distribution that is symmetric about the point $(1,0,0)_c$, i.e., $(1,\pi/2,0)_s$, and we will define a ``cut-off'' (range) of the Haar-uniform generation of the points on the curved surface, so that a 
(possibly small)
symmetric region around $(1,\pi/2,0)_s$ is utilized. 
Consider a plane parallel to the $y-z$ plane cutting the surface of the sphere in two parts. Let us denote the perpendicular distance 
of the plane from the point $(1,0,0)_c$ 
as $r_U$. 
This parameter, $r_U$, describes the range of the distribution and can vary within \([0,2]\), as the radius of the Bloch sphere is unity. The random numbers $(1,\theta,\phi)_s$ are drawn from the surface of that part of the sphere which belongs to the positive $x$ direction of the plane. 
Haar-uniform distribution on sphere can be constructed by uniform selection of $\cos{\theta}$ and $\phi$ within the range specified by range parameter $r_U$. Since the distribution is symmetric about the point $(1,0,0)_c$, whatever be the range, the direction of the mean of the distribution will always be $(1,0,0)_c$. 
For each value of $r_U$, the corresponding strength of disorder, $\sigma_U$, can be calculated using Eq.~\eqref{eq5}. In Fig.~\ref{fig1}, we plot 
the points chosen from the Haar-uniform distribution with different values of $\sigma_U$, on a purple sphere. We plot 
$10^4$ points for each value of $\sigma_U$. 

Later, we will evaluate scaling of the DQWs with the disorder in its coin being chosen from different Haar-uniform distributions. We will employ \(\sigma_U\) to quantify the strength of the disorder being used.

\subsection{Spherical normal distribution}
\label{subsec2}
Next we consider a ``spherical normal'' distribution, i.e., we
want to choose the random biased Hadamard gates such that the \((\theta,\phi)\) corresponding to 
the \(|\xi(\theta,\phi)\rangle\), of the gates,
are ``spherical normally'' distributed on the Bloch sphere, around the state $(\ket{0}+\ket{1})/\sqrt{2}$.
%
The 
von Mises - Fisher distribution  (vMFD) has been interpreted as a normal distribution on a sphere~\cite{ref73}. 
It is 
described using two parameters, $\kappa_N ,~\boldsymbol{\mu}_N$, 
and the probability measure is
\begin{equation*}
    p(\theta,\phi) d\theta d\phi= A
    \exp(\kappa_N\textbf{x}^T \cdot \boldsymbol{\mu}_N)\sin{\theta} d\theta d\phi,
\end{equation*}
where $\kappa_N \in [0, \infty)$ is called the concentration parameter and $\boldsymbol{\mu}_N\equiv(\mu_x,\mu_y,\mu_z)_c$ represents the mean direction.  Here, $\textbf{x}\equiv (\sin{\theta}\cos{\phi},\sin{\theta}\sin{\phi},\cos{\theta})_c$
and \(A\) is the normalization constant. As one increases the value of $\kappa_N$, the distribution becomes more and more concentrated around the mean direction, $\boldsymbol{\mu}_N$. The vMFD becomes the uniform distribution for $\kappa_N \to 0$, whereas it leads to the ordered case 
for $\kappa_N\rightarrow\infty$. 
Thus the strength of the disorder, $\sigma_N$, is a function of $\kappa_N$.

Considering a mean direction along $(0,0,1)_c$, we get the simpler-looking distribution,
given by
\begin{eqnarray*}
p(\theta) d\theta &=& \frac{\kappa_N}{2 \sinh {\kappa_N}} \exp({\kappa_N \cos{\theta}}) \sin{\theta} d\theta,\\p(\phi)d\phi&=&\frac{1}{2\pi}d\phi.
\end{eqnarray*}
To take advantage of this simpler form, we choose random points $(\theta,\phi)$ from this  distribution which has mean at $(0,0,1)_c$, and then rotate those points to get a distribution with mean direction $(1,0,0)_c$.

\begin{figure}[h!]
\includegraphics[scale=0.5]{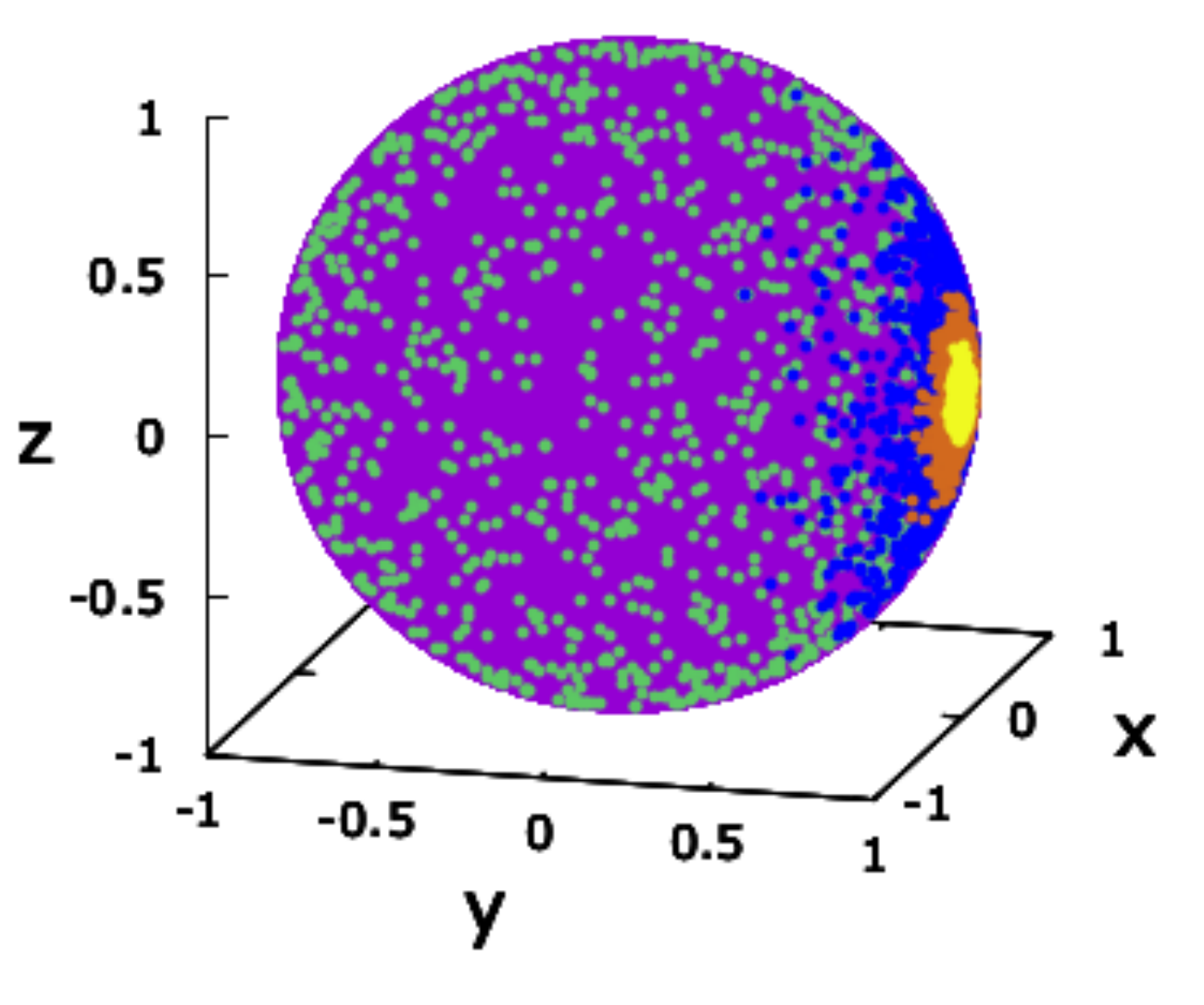}
\caption{Spherical normal distribution.
We plot points on a purple unit sphere, randomly generated from von Mises - Fisher distributions, with mean direction as 
$(1,0,0)_c$ and different concentration parameters, $\kappa_N$.
The different colors,
\emph{viz.} yellow, orange, blue, and green, denote points chosen from distributions with $\kappa_N=$ 500, 100, 10, and 0.1 
or $\sigma_N=$ 0.063, 0.142, 0.456, and 1.68
respectively. We have generated $10^3$ points for each concentration value.
All quantities used are dimensionless.} 
\label{fig2} 
\end{figure}

In Fig.~\ref{fig2}, we plot  points $(1,\theta,\phi)_s$ chosen randomly from  vMFDs with $\boldsymbol{\mu}_N= (1,0,0)_c$ and varying $\kappa_N$ or $\sigma_N$. One can see in Fig. \ref{fig2} that spreading of the distribution increases with $\sigma_N$.
The disorder averaged dispersion of the spread of the position of the quantum walker, when the disorder in the biased Hadamard gate is chosen from the vMFD distribution, is analyzed in  
%
Sec.~\ref{R2}.

\subsection{Circular distribution}
\label{subsec3}
Discrete probability distributions form an important category for studies of disorder incorporation in physical systems. 
If the disorder was inserted in a scalar quantity, the discrete probability distribution that 
would typically be considered is the one 
which takes certain values, \(\pm r_D\), assuming that the intended mean is at zero.
Arguably, the parallel in the case when we are required to scatter the distribution on a sphere, 
will be given by circles on the sphere, symmetrically placed around the intended mean direction. We note that when generalized in this way, the distribution is actually not discrete, but no discrete distribution will be rotationally symmetric about a mean direction.

\begin{figure}[h!]
\includegraphics[scale=0.5]{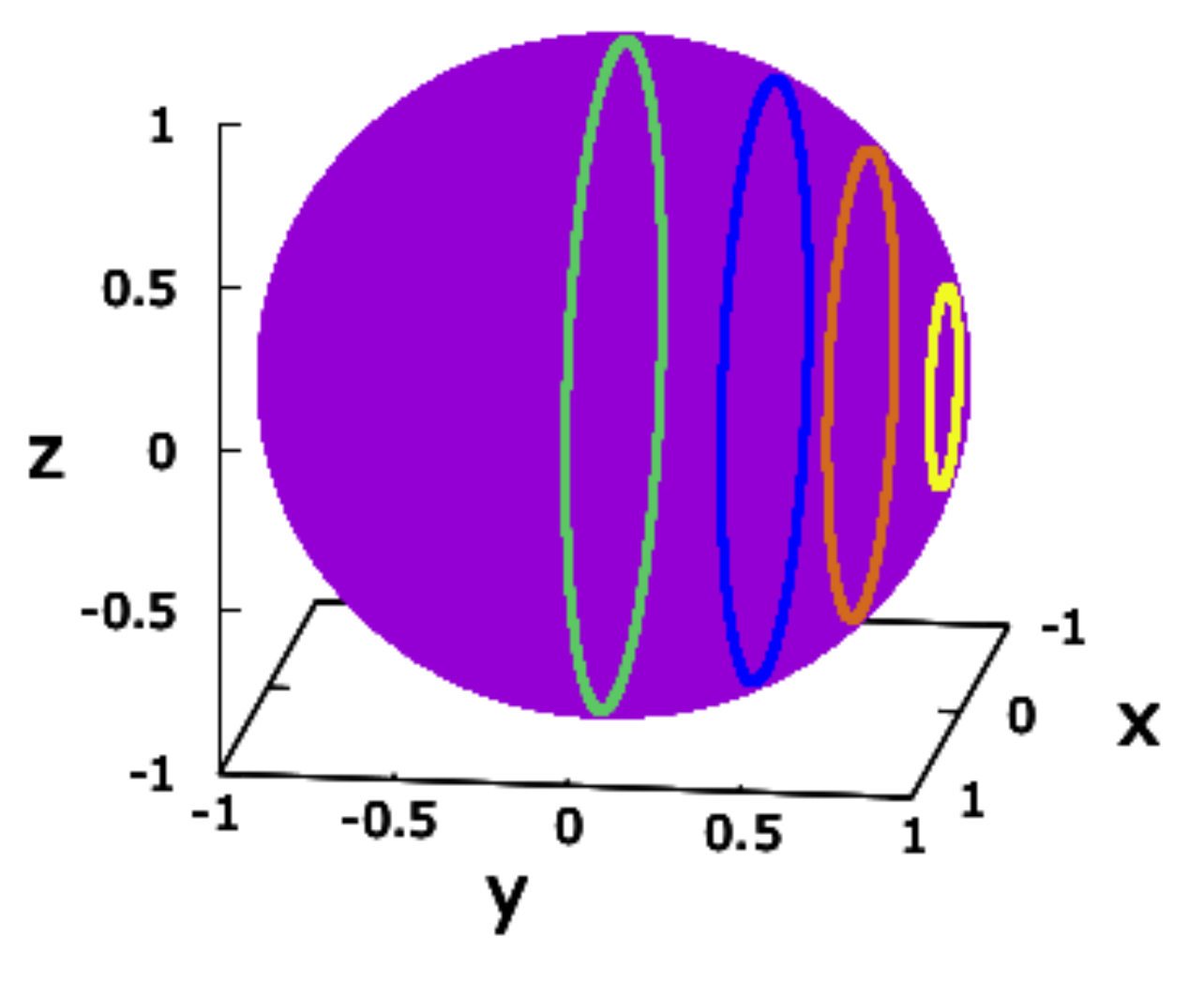}
\caption{Circular distribution on a  sphere. We generate, and plot, points chosen from circular distributions on the unit sphere of  different  \(\sigma_C\). For each range, we generate \(10^4\) points. 
The different colors, \emph{viz.} yellow, orange, blue, and green represent points drawn from distributions of strength  $\sigma_C= 0.305$, $0.775$, $1.12$, and $1.57$ respectively. The perpendicular distances of the planes containing the circularly distributed points from $(1,0,0)_c$ are  0.3 (yellow), 0.7 (orange), 0.9 (blue), and 1 (green) respectively. All quantities used are dimensionless.} 
\label{fig3} 
\end{figure}

Since we wish to have the mean direction as $(1,\pi/2,0)_s$, these circles will 
lie on planes  parallel to the $y-z$ plane. We name these as the circular distributions.
We will use the quantity  defined in Eq.~\eqref{eq5} to quantify the strength of the circular distribution, and denote it by $\sigma_C$.
Fig.~\ref{fig3} portrays  circular distributions for different values of $\sigma_C$. In the succeeding section, this distribution will be among the ones that are used for disorder insertion in the coin operator.




\subsection{Spherical Cauchy-Lorentz distributions}
\label{subsec4}

In studies about disorder insertion in system parameters on a flat 
parameter space, one often considers the Cauchy-Lorentz distribution, which, although has a profile similar to the Gaussian one, is starkly different from it, as the former does not have a finite mean and standard deviation, though a Cauchy mean value does exist.
In the literature, there are at least two distributions that has been claimed to be generalizations to
the sphere of the Cauchy-Lorentz distribution on a flat parameter space. 

\subsubsection{\textbf{Spherical Cauchy-Lorentz I}}
\label{sharmita}
The probability measure of one of them is given by
$$p(\theta,\phi) d\theta d\phi = \frac{1}{4\pi \ln{\delta}} \frac{(\delta^2 -1)}{(\delta^2 +1)-(\delta^2 -1)\textbf{x}^T \cdot \boldsymbol{\mu}_1} \sin{\theta}  d\theta d\phi.$$ 
It is dependent on the two parameters, $\delta$ and $\boldsymbol{\mu}_1$, acting respectively 
as
a concentration parameter and the mean direction.  The range of $\delta$ is $[0,\infty)\setminus \{1\}$. As $\delta \to 1$, the distribution becomes more and more uniform over the whole unit sphere~\cite{ref74}. 
The mean direction is \(\boldsymbol{\mu}_1\) for \(\delta >1\), and \(-\boldsymbol{\mu}_1\) for \(\delta <1\).

\begin{figure}[h!]
\includegraphics[scale=0.5]{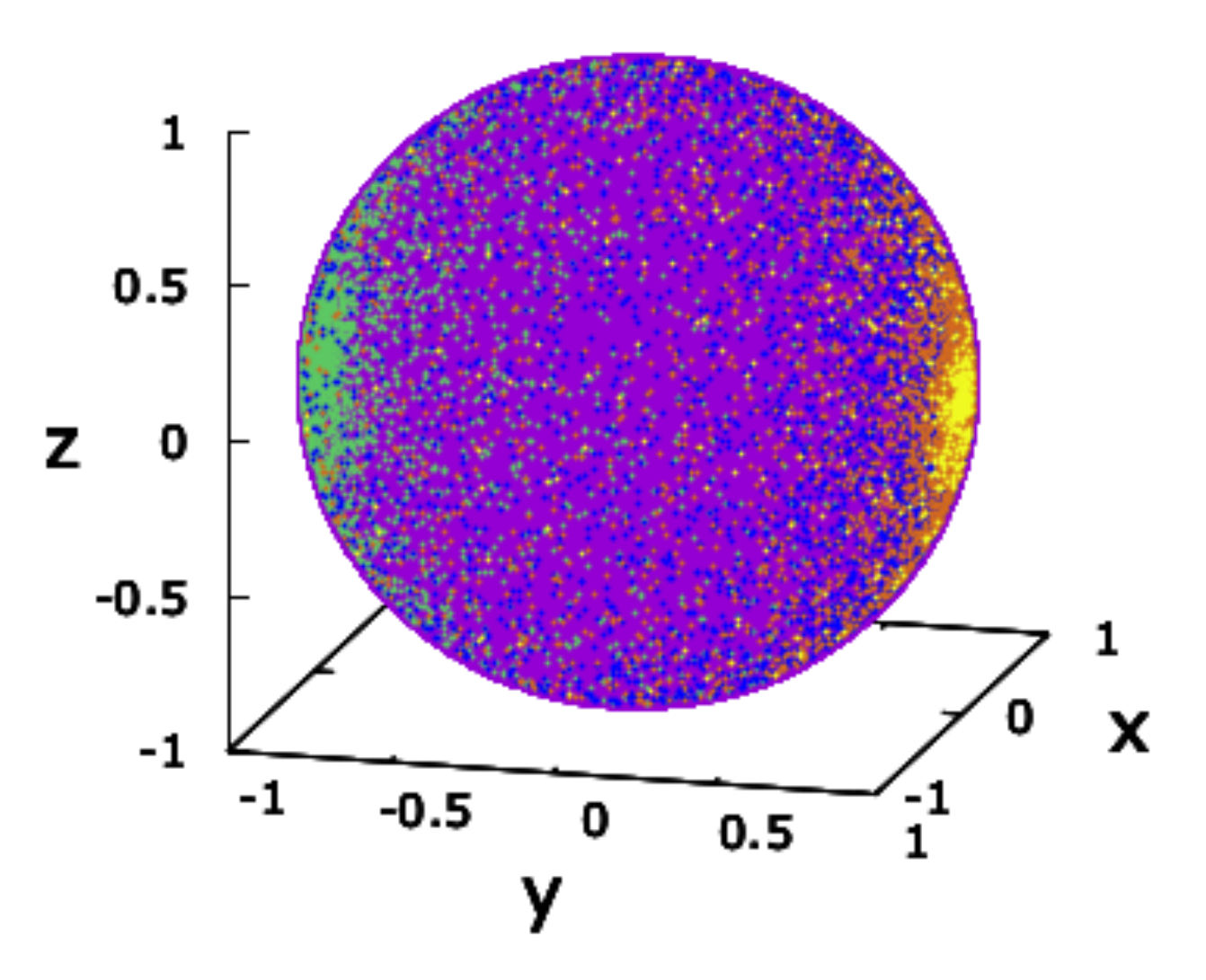}
\caption{Nature of Cauchy-Lorentz I distribution on a  sphere. We sample $10^3$ pairs of angles, $(\theta,\phi)$, from each of the Cauchy-Lorentz I distributions having $\delta = 10^4, 10^2, 10, 10^{-2}$. They are plotted on a unit purple sphere 
using yellow, orange, blue, and green  colors respectively. 
The corresponding values of $\sigma_{C1}$ are, respectively, 0.535, 0.756, 1.05, 2.74. 
All quantities used are dimensionless.} 
\label{fig4} 
\end{figure}

To generate points from this distribution, we first assume that the direction of the mean is along $z$-axis, i.e. $\boldsymbol{\mu}_1=(0,0,1)_c$. 
The measure for the distribution, therefore, becomes
\begin{eqnarray}
p(\theta) d\theta &=& \frac{1}{2 \ln{\delta}}  \frac{(\delta^2 -1)}{(\delta^2 +1)-(\delta^2 -1)\cos{\theta}} \sin{\theta} d\theta,\nonumber\\
p(\phi) d\phi&=& \frac{1}{2\pi}d\phi.\nonumber
\end{eqnarray}
After generating points using this  distribution, we rotate the points to get a distribution with mean along the $x$-axis. A few such distributions are exhibited in Fig.~\ref{fig4} 
for certain values of $\sigma_{C1}$,
the strength of the disorder. It is visible from Fig.~\ref{fig4} that as $\delta$ decreases from a higher value towards unity, the distribution becomes less concentrated along the mean direction and more uniform over the sphere. But if $\delta$ become less than 1 the distribution will drop its uniform nature and will again start to get concentrated, but now around the  direction opposite to that of $\boldsymbol{\mu}_1$. 

\subsubsection{\textbf{Spherical Cauchy-Lorentz II}}

Another distribution on the unit sphere that has been considered as a generalization of the Cauchy-Lorentz distribution can be defined again using two parameters, \emph{viz.} $\rho$ and $\boldsymbol{\mu}_2$, and for which the probability measure is given by
$$p(\theta,\phi) d\theta d\phi= \frac{1}{4\pi} \frac{(1-\rho^2)^2}{(1+\rho^2 - 2\rho \boldsymbol{\mu}_2^T \cdot \boldsymbol{x})^2}  \sin{\theta}  d\theta d\phi,$$
where $\rho \in \lbrack0,1)$ and $\boldsymbol{\mu}_2$ work as the concentration parameter and the  direction of the mode of the distribution respectively. As $\rho$ goes to zero, the distribution becomes uniform, and on the other hand, the distribution converges to a point distribution 
when $\rho$ tends to 1~\cite{ref75}.

\begin{figure}[h!]
\includegraphics[scale=0.5]{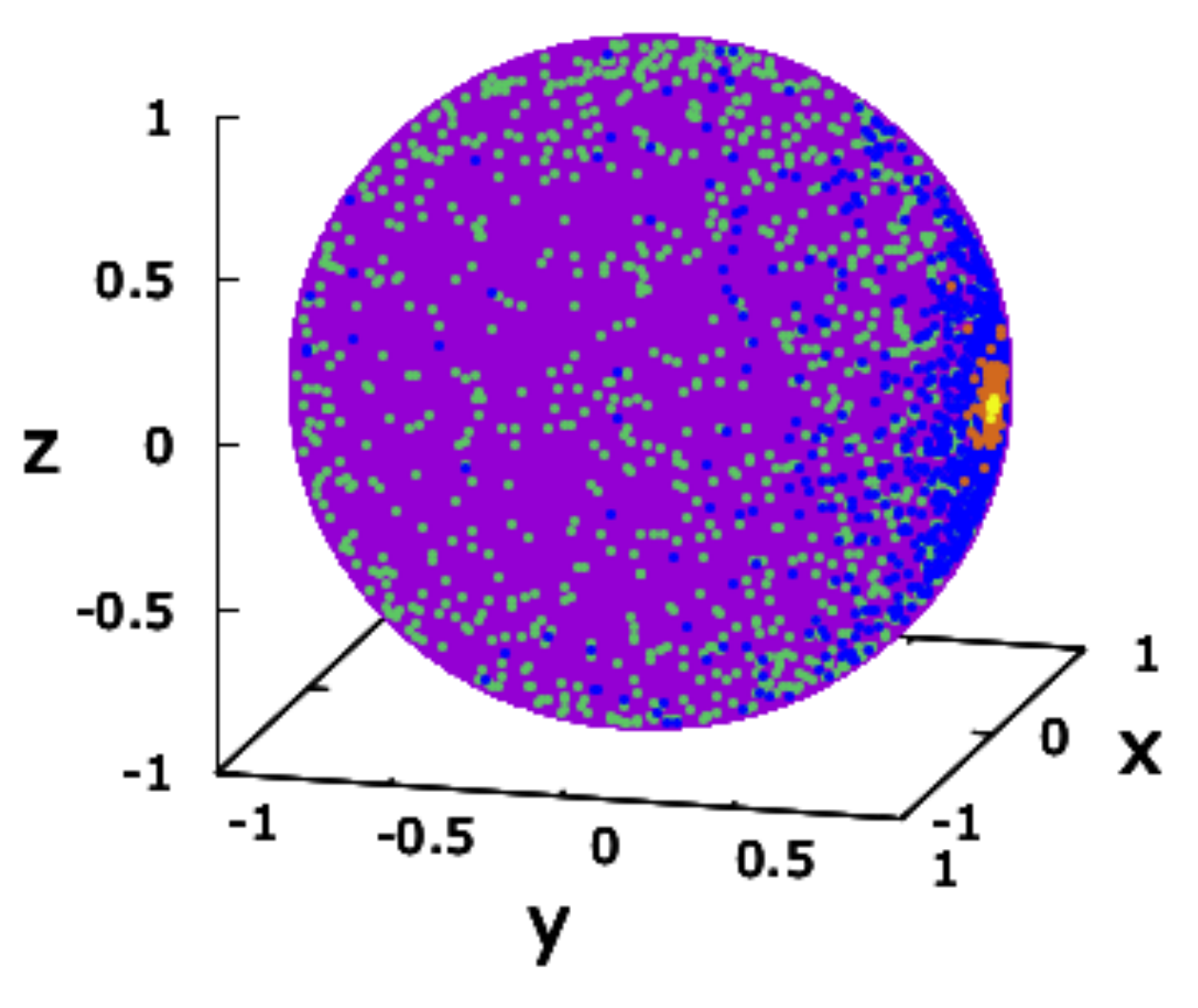}
\caption{Cauchy-Lorentz II distribution on a sphere. We generate, and plot, \(10^3\) points for every chosen \(\rho\) 
from the Cauchy-Lorentz II distribution on the purple unit sphere. 
The concentration parameter is chosen as 0.999, 0.98, 0.7, and 0.1, and represented using yellow, orange, blue, and green colored dots respectively. 
The corresponding amounts of disorder are found to be \(\sigma_{C2}\)= 0.00661, 0.0577, 0.626, and 1.57, respectively.
All quantities used are dimensionless.} 
\label{fig5} 
\end{figure}

Again taking the direction of the mode along the $z$-axis, the measure for the distribution reduces to
\begin{eqnarray*}
p(\theta) d\theta &=& \frac{1}{2} \frac{(1-\rho^2)^2}{(1+\rho^2 - 2\rho \cos \theta)^2 }  \sin{\theta} d\theta,\\
p(\phi)d\phi&=&\frac{1}{2\pi}d\phi.
\end{eqnarray*}
Generating points using this distribution, and then rotating the points appropriately, we can get the required distribution having mode along the $x$-axis. The variation of the pattern of the distribution 
can be seen in Fig.~\ref{fig5}.
The strength of the disorder is denoted as $\sigma_{C2}$.

The scaling of the disorder averaged dispersion of the quantum walk for both the spherical Cauchy-Lorentz distributions are analyzed in the succeeding section.


\section{Response to disorder}
\label{sec6}

The scaling exponent of  dispersion of an ordered Hadamard quantum walk is unity, being double of that for the classical random walker. But we will find, in this section, that in the presence of disorder in the coin operator, this ballistic spread in the quantum case becomes sub-ballistic, although remains super-diffusive, with the diffusive spread being reached by the classical random walker.
While this qualitative feature remains unchanged for all the disorder distributions that we have considered, the actual scale exponents and their behavior with strength of the disorder differ significantly from one distribution to the other.

Here we estimate  scalings of the dispersions of  disordered quantum walks for each of the distributions discussed in the preceding section, for varying disorder strengths. 
%
After fixing a distribution of the disorder, we analyze the scaling exponent, $\alpha$, of the quantum walker's spread. In particular, we look at its behavior as a function of the strength of the disorder. 
In each of these cases, 
we try to fit a  curve, and estimate the curve parameters via the least-squares method. We also estimate the  fitting error in each case.
The method is briefly defined here for completeness. For a set of \(N\) data points, \((x_1,y_1),(x_2,y_2),\ldots,(x_N,y_N)\), to which we fit the curve \(y=y(x|a_1,a_2,\ldots,a_M)\), involving \(M\) parameters, \(a_1,~a_2,\ldots,~a_M\), 
we set
\[\chi^2 = \sum_{i} \left[ \frac{y_{i} - y(x_{i} |a_{1},\ldots,a_{M})}{\tilde{\sigma}_{i}} \right]^2.\]
The quantity \(\tilde{\sigma}_i\) represents the standard deviation of the \(i^{\text{th}}\) data point. We assume that they are equal to each other in our analysis. 
The maximum-likelihood estimate of the fitting parameters are obtained by minimizing \(\chi^2\), to obtain \(\chi^2_{\min}\). The corresponding error is given by 
\(\sqrt{\tilde{\sigma}^2{\chi^2_{\min}} /(N-M)}\),
where \(\tilde{\sigma}\) is the common value of the \(\tilde{\sigma}_i\).

We begin by going over to the momentum representation, which aids in the evaluation of the scaling exponents. Subsequently, we consider the different disorder distributions and their respective exponents.  

\subsection{Momentum representation}
\label{sec4}



The amplitude of the quantum walker at the \(n^{\text{th}}\) position at time $t$ can be written as
\[  \ket{\Psi(n,t)} = \left (\begin{array}{c}
\Psi_{1}(n,t)   \\
\Psi_{0} (n,t) \\
\end{array} \right),\] 
where $\Psi_{1}(n,t)$ and $\Psi_{0}(n,t)$ denote the amplitudes of the quantum walker at position $n$ at time $t$, with the quantum coin being in $\ket{1}$ and $\ket{0}$  respectively.
%
To move on to the next step, we have to first operate the biased Hadamard gate on the coin's space and then have to act the shift operator on the composite space.
The coin operator, \(C\), is now to be chosen as the 
biased
Hadamard gate,  given by   
\begin{equation}
\label{shapla}
    H(\theta,\phi)=\left(
    \begin{matrix}
    \cos{(\theta/2)}&&\sin(\theta/2)\\\sin(\theta/2)e^{i\phi}&&-\cos{(\theta/2)e^{i\phi}}
    \end{matrix}
    \right),
    \end{equation}
where $\theta$ and $\phi$ are randomly chosen parameters from the probability distribution of the disorder.

Since the shift operator moves the walker only one step towards 
its left or right, the amplitude $\ket{\Psi (n,t+1)}$ will have contributions from $\ket{\Psi (n-1,t)}$ and $\ket{\Psi (n+1,t)}$ only. Hence the former can be expressed in terms of the latter in the following way 
(compare with~\cite{ref2}):
\begin{equation}
\ket{\Psi (n,t+1)} =  M_{-} (\theta,\phi) \ket{\Psi (n-1,t)} +  M_{+}(\theta,\phi) \ket{\Psi (n+1,t)}.
\label{eq1}
\end{equation}
It can be seen 
that 
\begin{eqnarray*}
M_-(\theta,\phi)&=&\left(\begin{matrix}
0 & 0  \\
\sin(\theta/2)e^{i\phi}&-\cos{(\theta/2)e^{i\phi}} \\
\end{matrix}\right),\\
M_+(\theta,\phi)&=&\left(\begin{matrix}\cos{(\theta/2)}&\sin(\theta/2)\\
0 & 0\\
\end{matrix}\right).
    \end{eqnarray*}
Eq.~\eqref{eq1} is the recursion relation for the quantum walk using the disordered Hadamard gate. The Fourier transform of $\ket{\Psi (n,t)}$ 
is
\begin{eqnarray*}
\ket{\tilde{\Psi}(k,t)} = \sum_{n=-\infty}^{+\infty}  \ket{\Psi(n,t)} e^{ikn}. \end{eqnarray*}
Then using Eq.~\eqref{eq1}, we can write $\ket{ \tilde{\Psi}(k,t+1)}$ in terms of $ \ket{\tilde{\Psi}(k,t)}$ in the following manner:
\begin{eqnarray*}
 \ket{\tilde{\Psi}(k,t+1)} &=& \sum_{n=-\infty}^{+\infty} (M_{-} \ket{\Psi(n-1,t)} + M_{+} \ket{\Psi(n+1,t)}) e^{ikn}, \\
\Rightarrow\ket{\tilde{\Psi}(k,t+1)} &=& (e^{ik} M_{-} + e^{-ik} M_{+})\ket{ \tilde{\Psi}(k,t)}\\
\Rightarrow\ket{\tilde{\Psi}(k,t+1)} &=& M_{k}(\theta,\phi) \ket{ \tilde{\Psi}(k,t)}.
\end{eqnarray*}
Here, $M_{k}(\theta,\phi)$ is given by
\begin{equation*}
 M_k(\theta,\phi) = \left(
    \begin{matrix}
    e^{-ik}\cos{(\theta/2)}&&e^{-ik}\sin(\theta/2)\\e^{ik}\sin(\theta/2)e^{i\phi}&&-e^{ik}\cos{(\theta/2)e^{i\phi}} \label{eq2}
    \end{matrix}
    \right).
\end{equation*}
Therefore, by starting with the initial state $\ket{\tilde{\Psi}(k,0)}$, if we take $t$  steps, then the final momentum-space amplitude is given by
\begin{equation*}
\ket{\tilde{\Psi}(k,t)} = M_k(\theta_t,\phi_t) M_k(\theta_{t-1},\phi_{t-1})....... M_k(\theta_1,\phi_1) \ket{ \tilde{\Psi}(k,0)}.
\end{equation*}
In the noiseless scenario, i.e., in the case when there is no disorder, the above equation reduces to $\ket{\tilde{\Psi}(k,t)} = M_k^t(\pi/2,0)\ket{ \tilde{\Psi}(k,0)}$. This final state, and the corresponding spread in position, can be obtained by 
using the spectral decomposition of $M_k(\pi/2,0)$, and subsequently finding the inverse Fourier transform of $\ket{\tilde{\Psi}(k,t)}$. In the disordered case, however, 
%
$M_k(\theta,\phi)$ 
is 
different 
in each step.


\subsection{Uniform disorder}
\begin{figure}[h!]
\includegraphics[scale=.7]{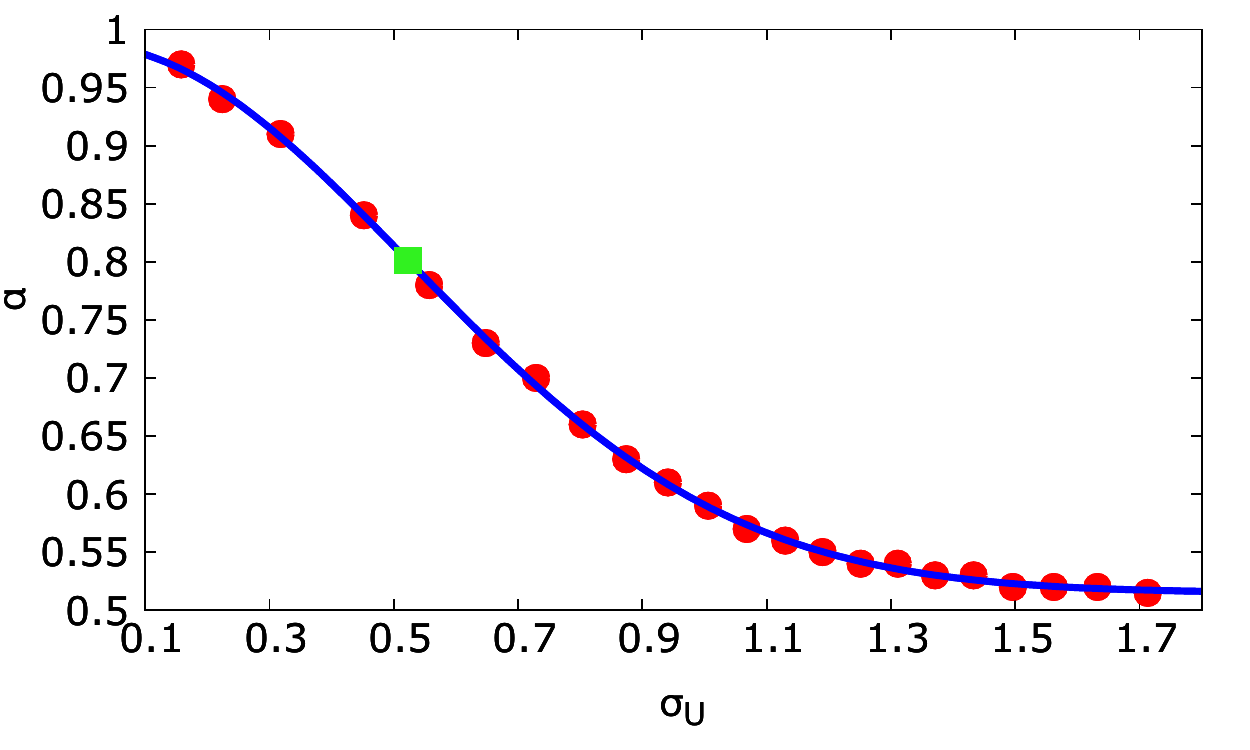}
\caption{Response of spread of quantum walker to uniform disorder. We present here the behavior of the scaling exponent of disorder averaged dispersion - as quantified by the disorder averaged standard deviation - of  the quantum walker in presence of the uniform glassy disorder in the coin operator. 
We plot the scaling exponent \((\alpha)\) against the strength, \(\sigma_U\), of the uniform disorder applied. 
The dots represent the values of \(\sigma_U\) where a numerical analysis is performed, and the corresponding scaling exponent is explicitly found. The joining curve is the fitted curve, via the method of least squares. See text for further details. All quantities used are dimensionless. 
} 
\label{fig6} 
\end{figure}

We now consider the case of uniform disorder with an arbitrary but fixed 
strength. Since the parameter space is not flat, one needs to consider the suitable Haar-uniform probability measure, as discussed in Sec.~\ref{subsec1}. 
For a given value of \(\sigma_U\), of the uniform disorder distribution, we determine the 
standard deviation of the walker's position distribution for each disorder configuration.
We subsequently find the average of the standard deviations
over the different disorder configurations, each of which is picked independently and at random from the uniform distribution on the sphere for 
the chosen strength, \(\sigma_U\). Subsequently, we calculate the scaling exponent (see Eq.~(\ref{ekTa-gaan-likho})) for 
the chosen \(\sigma_U\). We repeat the calculation to check for convergence by choosing a larger number of disorder configurations for the same \(\sigma_U\). We perform this arithmetic for several values of \(\sigma_U\), to understand the profile of the scaling exponent as a function of the disorder strength, \(\sigma_U\). 
%
%
%
%
In Fig.~\ref{fig6}, 
we plot the scaling exponent, \(\alpha\), versus the 
disorder strength, \(\sigma_U\). The profile matches well with a Gaussian function:
 \begin{equation}
     \alpha = a_U \exp(-b_U \sigma_U^2) +c_U, \label{eq7}
 \end{equation}
 where the the parameter values are given by $a_U = 0.472 \pm 0.00223$, $b_U = 1.83 \pm 0.0227$, 
 and $c_U = 0.515 \pm 0.00140$, being obtained from a least-square fitting method, with the error being $0.00321$. 
 The numbers appearing after the \(\pm\) sign indicate the 95\% confidence interval.
 All numerical data are correct to three significant figures. 
 %
 We observe that in response to uniform disorder, 
 the scaling exponent of dispersion of the quantum walker's position reduces from the unit scaling exponent of the ordered case.
 However, it never reaches the value 0.5, and even when the disorder 
 strength reaches its maximum value, 1.71, the scaling exponent is strictly higher than 1/2, the classical random walker's position distribution scaling exponent. 
 The dependence of $\alpha$ on $\sigma_U$ shows a concave nature for weak disorder and convex nature for high disorder. The transition point from concave to convex is shown in Fig. \ref{fig6}, using a green square point at 0.523.
 

\subsection{Spherical normal disorder}\label{R2}
\begin{figure}[h!]
\includegraphics[scale=.7]{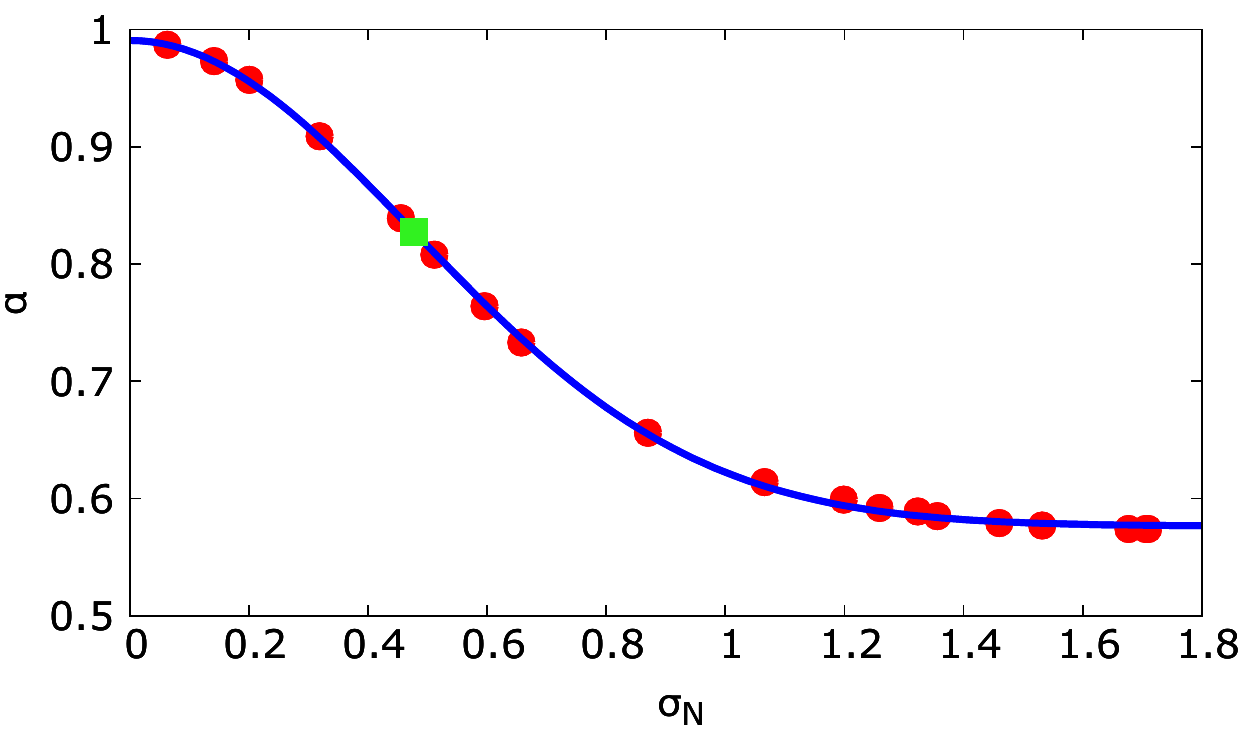}
\caption{Scaling behavior in presence of spherical normal disorder. 
We plot the strength of the von Mises - Fisher distribution, $\sigma_N$, along the horizontal axis and the corresponding value of scaling, $\alpha$, along the vertical axis. The dots represent the numerical values obtained, while the line is the fitted curve.  
All quantities used are dimensionless. 
The point where the profile changes its behavior from being concave to convex, is marked with a green square. It is at \(\sigma_N \approx 0.477\).} 
\label{fig8} 
\end{figure}

We now analyze the response of the scaling exponent to the disorder that follows the von Mises - Fisher distribution. 
The profile of \(\alpha\) against 
the disorder strength quantifier, $\sigma_N$, is exhibited 
in Fig.~\ref{fig8}. For increasing  disorder strength, there is a crossover from a concave to a convex function, which we depict in the figure using a green square. 
Like in the case of uniform disorder, the spherical normal distribution  also has  a \(e^{-(\cdot)^2}\) fit, for the scaling exponent against the disorder strength.


From Fig~\ref{fig8}, it can be seen that the scaling exponent of the quantum walk decreases with $\sigma_N$ up to a certain value, after which it becomes almost a constant. For large \(\sigma_N\), the exponent converges to a value that is higher than the same in the classical case, and also a little higher than that for large uniform disorder.


The full range of the data, which is presented in Fig.~\ref{fig8}, can be fitted with
\begin{equation}
    \alpha = a_N\exp(-b_N\sigma_N^2)+c_N,\label{eq3}
\end{equation}
for $a_N = 0.414 \pm 0.00176$, $b_N = 2.20 \pm 0.0271$, and $c_N = 0.576 \pm 0.00104$.
The least-square fitting error is 
0.00275. 

\subsection{Circular disorder}
\begin{figure}[h!]
\includegraphics[scale=.7]{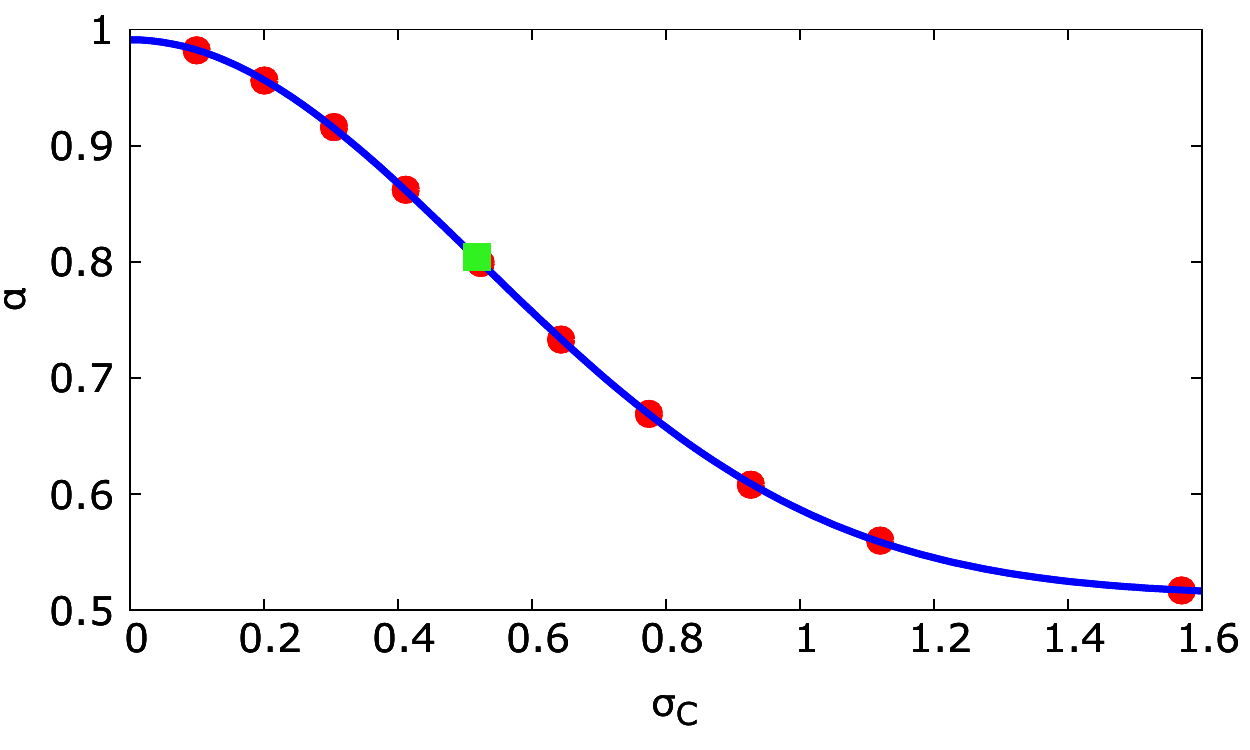}
\caption{How does circular disorder present in coin affect the motion of a quantum walker? 
The considerations are the same as in the preceding two figures, except that the disorder inflicted is from the circular distribution, and the horizontal axis represents the circular disorder strength, \(\sigma_C\). The green square indicates the transition of the curve from concave to convex behavior. 
See text for more details.
All quantities used  are dimensionless.
} 
\label{fig9} 
\end{figure}

When the disorder present in the coin operator of  quantum walk follows the circular distribution, then the scaling of the DQWs varies with the strength $\sigma_C$ of the distribution as in Fig.~\ref{fig9}, and the fitting function is
\begin{equation}
\alpha = a_C\exp(-b_C\sigma_C^2)+c_C,\label{eq8}  
\end{equation}
where $a_C = 0.479 \pm 0.000860$, $b_C = 1.86 \pm 0.00897$, and $c_C = 0.513 \pm 0.000810$. The corresponding error is approximately \(= 0.000846\). Though the dependence of $\alpha$ on $\sigma_C$ appears to be Gaussian, the curve has a reflection symmetry around \(\sigma_C=\pi/2\). 
If $\sigma_C$ is increased to values higher than $\pi/2$, i.e, if the circle of the disorder moves closer to the point $(-1,0,0)_c$ than $(1,0,0)$, then the spreading of the quantum walker's position will again start to increase. The reason behind this 
behaviour is explained in Sec.~\ref{C1}, which deals with a distribution having a similar feature.
Like for the vMF and uniform distributions, the scaling exponent is concave for weak disorder, which then  switches to a convex behavior for higher values of the disorder strength. The point from where it switches to convex from concave is approximately 
$\sigma_C= 0.518$. This point is denoted by a green square in  Fig.~\ref{fig9}.
Like both the previous types of disorder, the minimum value of the scaling exponent is a little higher than the classical value (of 1/2). 


\subsection{Spherical Cauchy-Lorentz disorder}

\subsubsection{\textbf{Cauchy-Lorentz I}}
\label{C1}
\begin{figure}[h!]
\includegraphics[scale=.7]{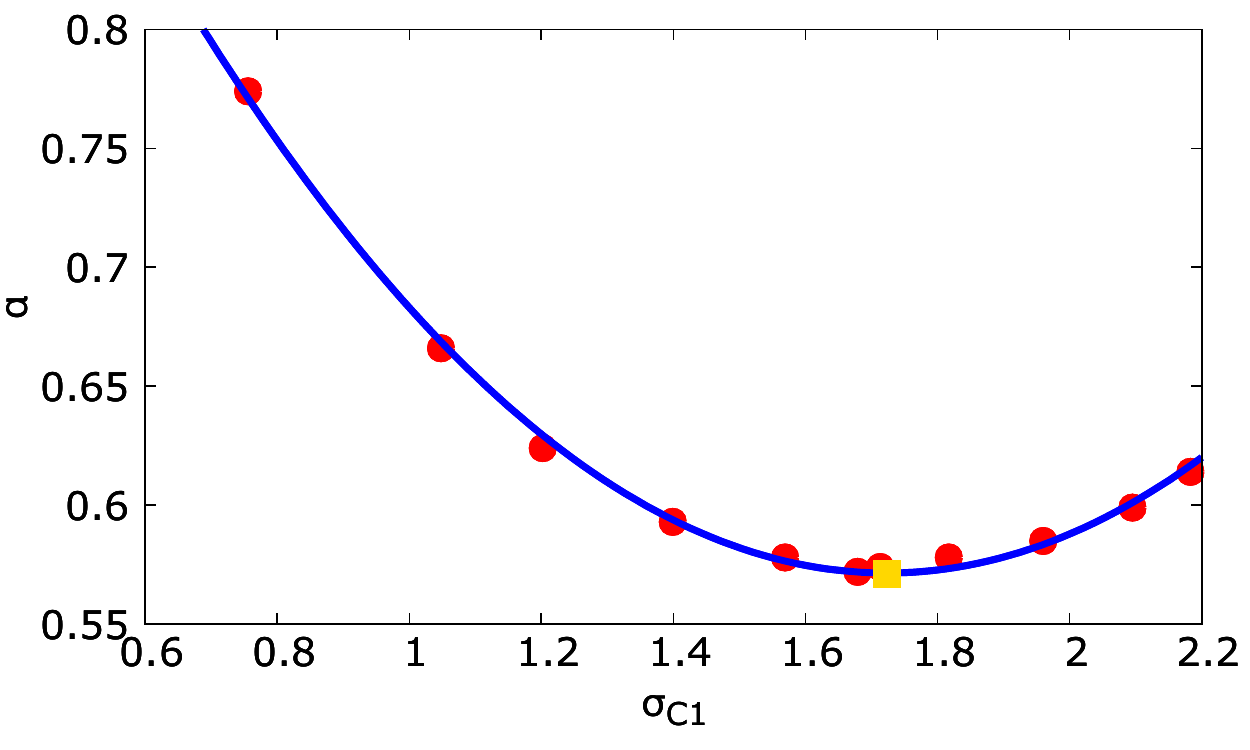}
\caption{Illustration of scaling in presence of Cauchy-Lorentz I disorder in DQWs. 
The considerations are the same as in the preceding few figures, except that the disorder distribution is Cauchy-Lorentz I, and the horizontal axis represents the strength of that disorder, \emph{viz.} $\sigma_{C1}$. All quantities used are dimensionless.
} 
\label{fig10} 
\end{figure}

The scaling of disorder averaged dispersion of the quantum walk, with respect to  strength of the disorder, distributed as Cauchy-Lorentz I, is presented in Fig.~\ref{fig10}. The data can be fitted 
with the  parabolic function,
\begin{equation}
     \alpha = a_1 \sigma_{C1}^2 + b_1\sigma_{C1} + c_1, \label{eq15}
\end{equation}
where $a_1 = 0.213 \pm 0.00415$, $b_1 = -0.736 \pm 0.0118$, and $c_1 = 1.20 \pm 0.00773$. The corresponding least squares fitting error is 0.00316. 




The scaling exponent of the disorder averaged standard deviation of the walker's position distribution 
is minimal for $\delta \rightarrow 1$, which we denote using a yellow square point in Fig.~\ref{fig10}. When $\delta$ is further decreased, i.e., the $\sigma_{C1}$ is increased, the spreading starts to increase. Though the amount of disorder present in the system increases with $\sigma_{C1}$, its effect starts to decrease when $\sigma_{C1}$ crosses the value 1.71, because for higher values of $\sigma_{C1}$, the states $\frac{\ket{0}+\ket{1}}{\sqrt{2}}$ and $\frac{\ket{0}-\ket{1}}{\sqrt{2}}$ exchange their roles. [The cross-over point is at 1.72 if calculated by finding the minimum of the fitted curve, as in Fig.~\ref{fig10}.] Similar situations arises in case of circular disorder for $\sigma_C>\pi/2$. For example, in case of circular disorder, when $\sigma_C$ reaches its maximum value, i.e. $\pi$, the disordered Hadamard gate, instead of mapping the state $\ket{0}$ to $\frac{\ket{0}+\ket{1}}{\sqrt{2}}$,  maps $\ket{0}$ to $\frac{\ket{0}-\ket{1}}{\sqrt{2}}$.


\begin{figure}[t!]
\includegraphics[scale=.7]{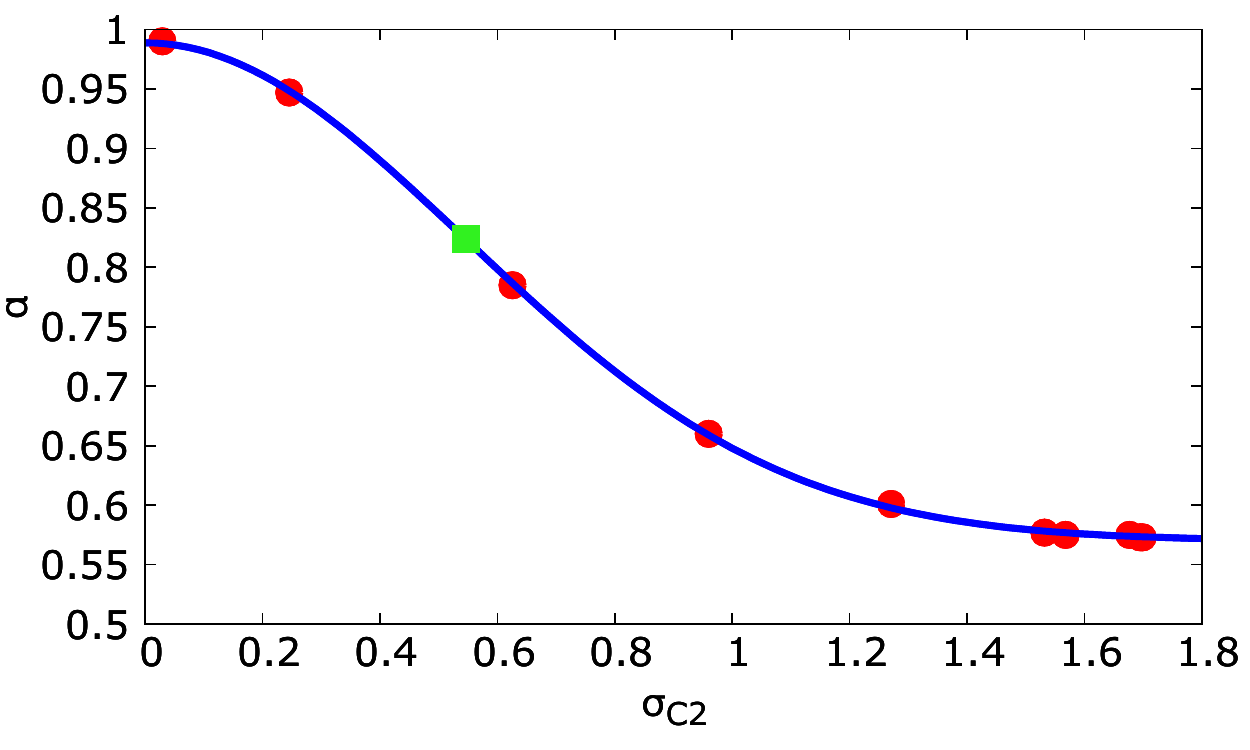}
\caption{Scaling behavior of dispersion due to disordered quantum coin following the  Cauchy-Lorentz II distribution. The considerations are the same as in the preceding few figures, except that the disorder distribution is Cauchy-Lorentz II, and the horizontal axis is for the strength of that disorder.
All quantities used are dimensionless.}
\label{fig11} 
\end{figure}

\subsubsection{\textbf{Cauchy-Lorentz II}}

We now calculate the $\alpha$ in presence of Cauchy-Lorentz II disorder in the quantum coin. The results are presented in Fig.~\ref{fig11}, in an \(\alpha\) versus disorder strength plot. We fit the data with the relation,
\begin{equation}
  \alpha = a_2 \exp(-b_2 \sigma_{C2}^2) + c_2,  \label{eq6}
\end{equation}
where $a_2 = 0.419 \pm 0.00166$, $b_2 = 1.68 \pm 0.0212$ and $c_2 = 0.570 \pm 0.000983$. The corresponding least squares error is \(0.00194\). 
The behaviour of the spreading is similar to that for the uniform and spherical normal disorders, i.e., the scaling exponent falls to a near but higher than classical value, with a transition from concave to convex. The inflection point is indicated using a green square in Fig.~\ref{fig11}, which occurs at $\sigma_{C2}=0.546$. 



\section{CONCLUSION}
\label{sec7}
In a real experiment, it is almost impossible to completely remove the presence of disorder in system parameters. Furthermore, disorder often induces advantages in system characteristics and resource efficiencies over the ordered case, and may uncover interesting phases of a system. It is therefore interesting to consider the response to disorder in system parameters of a quantum device and its advantage over the classical parallels. It is also interesting to note that disorder can now be artificially incorporated in physical systems realized in the laboratories. 

In this work, we analyzed the response of  spread of a quantum walker when the coin operation of the device's set-up is affected by ``glassy'' (or ``quenched'') disorder, that is, a type of disorder in which the equilibration time of the disorder is several orders of magnitude higher than the typical observation times that we are interested in. We considered several paradigmatic disorder distributions on the sphere, to consider disorder insertion in the quantum coin's ``disordered Hadamard operation''. We investigated in particular the scaling exponent of the disorder averaged spread of the quantum walker's position, as quantified by the standard deviation of the walker's position distribution. We found that the scaling exponent is such that the spread in the disordered case, for the distributions considered, is reduced from the ballistic spread in the ordered quantum case. However, the spread remains higher than the dispersive spread of the classical random walker. 
Most of the disorder distributions lead to slow, Gaussian-like reductions - with increasing disorder strength - of the scaling exponent of the spread, while for only one type distribution (a so-called ``spherical Cauchy-Lorentz'' distribution) the decay is parabolic. For most disorder distributions, therefore, the response to weak disorder is weak, with an inflection at a certain mid-range disorder strength, after which the response is strong.



\begin{acknowledgments}
We acknowledge the cluster facility of the Harish-Chandra
Research Institute for the numerical computations performed therein.
We acknowledge partial support from the Department of Science and Technology, Government of India through the QuEST grant (grant number DST/ICPS/QUST/Theme-3/2019/120). 
\end{acknowledgments}


\begin{thebibliography}{9}
\bibitem{ref1}
Y. Aharonov, L. Davidovich, and N. Zagury, Quantum random walks, Phys. Rev. A \textbf{48}, 1687 (1993).
\bibitem{ref2}
A. Nayak and A. Vishwanath, Quantum walk on the line, arXiv:quant-ph/0010117.
\bibitem{ref3}
A. Ambainis, E. Bach, A. Nayak, A. Vishwanath, and J. Watrous, One-dimensional quantum walks, Proceedings of STOC’01, pp. 37-49.
\bibitem{ref4}
J. Kempe, Quantum random walks - an introductory overview, Contemp. Phys. \textbf{44}, 307 (2003).
\bibitem{ref140}
V. Kendon, Decoherence in quantum walks - a review, Math. Struct. in Comp. Sci \textbf{17}, 1169 (2006).
\bibitem{add6}
T. Kitagawa, Topological phenomena in quantum walks; elementary introduction to the physics of topological phases, arXiv:1112.1882.
\bibitem{ref5}
O. M{\" u}lken and A. Blumen, Continuous-time quantum walks: models for coherent transport on complex networks, Phys. Rep. \textbf{502}, 37 (2011).
\bibitem{ref6}
T. Kitagawa, Topological phenomena in quantum walks: elementary introduction to the physics of topological phases,  Quantum Inf. Process. \textbf{11}, 1107 (2012).
\bibitem{ref7}
S. E. Venegas-Andraca, Quantum walks: a comprehensive review, Quantum Inf. Process. \textbf{11}, 1015 (2012).
\bibitem{ref8}
Y. Shikano, From discrete time quantum walk to continuous time quantum walk in limit distribution, J. Comput. Theor. Nanosci. \textbf{10}, 1558 (2013).

\bibitem{ref9}
F. Xia, J. Liu, H. Nie, Y. Fu, L. Wan, and X. Kong, Random walks: A review of algorithms and applications, IEEE Transactions on Emerging Topics in Computational Intelligence, \textbf{4}, 95 (2020).


\bibitem{ref12}
B. C. Travaglione and G. J. Milburn, Implementing the quantum random walk, Phys. Rev. A \textbf{65}, 032310 (2002).

\bibitem{ref11}
  A. M. Childs, R. Cleve, E. Deotto, E. Farhi, S. Gutmann, and D. A. Spielman, Exponential algorithmic speedup by quantum walk, Proc. 35th ACM Symposium on Theory of Computing (STOC 2003), pp. 59-68.




 \bibitem{ref13}
W. Dai, n-Qubit operations on sphere and queueing scaling limits for programmable quantum computer, arXiv:2109.14266.
\bibitem{ref10}
S. Akmal and C. Jin, Near-optimal quantum algorithms for string problems, arXiv:2110.09696.
 \bibitem{ref14}
 C. M. Chandrashekar and T. Busch, Localized quantum walks as secured quantum memory, Europhys. Lett. \textbf{110}, 10005 (2015).
\bibitem{ref15}
R. Asaka, K. Sakai, and R. Yahagi, Quantum random access memory via quantum walk, Quantum Sci. Technol. \textbf{6}, 035004 (2021).


\bibitem{ref17}
D. Aharonov, A. Ambainis, J. Kempe, and U. Vazirani, Quantum walks on graphs, arXiv:quant-ph/0012090.
\bibitem{ref16}
A. M. Childs, E. Farhi, and S. Gutmann, An example of the difference between quantum and classical random walks, Quantum Inf. Processing  \textbf{1}, 35 (2002).

\bibitem{ref18}
P. Kurzy\'{n}ski and A. W\'{o}jcik, Discrete-time quantum walk approach to state transfer, Phys. Rev. A \textbf{83}, 062315 (2011).
\bibitem{ref19}
 X. Zhan, H. Qin, Z.-h. Bian, J. Li, and P. Xue, Perfect state transfer and efficient quantum routing: a discrete-time quantum walk approach, 	Phys. Rev. A \textbf{90}, 012331 (2014).
 \bibitem{ref20}
\.{I}. Yal\c{c}{\i}nkaya and Z. Gedik, Qubit state transfer via discrete-time quantum walks, J. Phys. A: Math. Theor. \textbf{48}, 225302 (2015).
\bibitem{ref21}
H. Li, J. Li, and X. Chen, Discrete-time quantum walk approach to high-dimensional quantum state transfer and quantum routing, arXiv:2108.04923.
\bibitem{ref22}
M. Annabestani, M. Hassani, D. Tamascelli, and M. G. A. Paris, Multi-parameter quantum metrology with discrete-time quantum walks, arXiv:2110.02032.
\bibitem{ref23}
J. Mulherkar, R. Rajdeepak, and V. Sunitha, Quantum simulation of perfect state transfer on weighted cubelike graphs, arXiv:2111.00226.


\bibitem{ref35}
 C. A. Ryan, M. Laforest, J. C. Boileau, and R. Laflamme, Experimental implementation of a discrete-time quantum random walk on an NMR quantum-information processor, Phys. Rev. A \textbf{72}, 062317 (2005).
\bibitem{ref37}
H. B. Perets, Y. Lahini, F. Pozzi, M. Sorel, R. Morandotti, and Y. Silberberg, Realization of quantum walks with negligible decoherence in waveguide lattices, Phys. Rev. Lett. \textbf{100}, 170506 (2008).
\bibitem{ref36}
M. Karski, L. Frster, J. M. Choi, A. Steffen, W. Alt,
D. Meschede, and A. Widera, Quantum walk in position space with single optically trapped atoms, Science \textbf{325}, 174 (2009).
\bibitem{add1}
 O. Boada, L. Novo, F. Sciarrino, and Y. Omar, Quantum walks in synthetic gauge fields with three-dimensional integrated photonics, Phys. Rev. A \textbf{95}, 013830 (2017).
\bibitem{ref34}
Q.-Q. Wang, X.-Y. Xu, W.-W. Pan, K. Sun, J.-S. Xu, G. Chen, Y.-J. Han, C.-F. Li, and G.-C. Guo, Dynamic-disorder-induced enhancement of entanglement in photonic quantum walks, Optica \textbf{5}, 1136 (2018).
 \bibitem{ref33}
D. T. Nguyen, T. A. Nguyen, R. Khrapko, D. A. Nolan, and N. F. Borrelli, Quantum Walks in periodic and quasiperiodic Fibonacci fibers, arXiv:1911.01389.

\bibitem{ref32}
M. Gong, S. Wang, and C. Zha, Quantum walks on a programmable two-dimensional 62-qubit superconducting processor, Science \textbf{372}, 948 (2021).
\bibitem{add2}
O. L. Acevedo and T. Gobron, Quantum walks on Cayley graphs, J. Phys. A: Math. Gen. \textbf{39}, 585 (2006).
\bibitem{ref39}
G. Leung, P. Knott, J. Bailey, and V. Kendon, Coined quantum walks on percolation graphs, New J. Phys. \textbf{12}, 123018 (2010).
\bibitem{ref38}
R. Portugal, R. A. M. Santos, T. D. Fernandes, and D. N. Gonçalves, The staggered quantum walk model, Quantum Inf. Processing, \textbf{15}, 85 (2016).
\bibitem{ref41}
J. Jayakumar, S. Das, A. Sen(De), and U. Sen, Interference-induced localization in quantum random walk on clean cyclic graph, EPL \textbf{128}, 20007 (2019).





\bibitem{add3}
M. Kieferova and D. Nagaj, Quantum walks on necklaces and mixing, International Journal of Quantum Information \textbf{10}, 1250025 (2012).
\bibitem{ref44}
S. R. Jackson, T. J. Khoo, and F. W. Strauch, Quantum walks on trees with disorder: decay, diffusion, and localization, Phys. Rev. A \textbf{86}, 022335 (2012).

\bibitem{add4}
C. Benedetti, M. A. C. Rossi, and M. G. A. Paris, Continuous-time quantum walks on dynamical percolation graphs, EPL \textbf{124}, 60001 (2018).
\bibitem{add5}
P. Sin and J. Sorci, Continuous-time quantum walks on Cayley graphs of extraspecial groups, arXiv:2011.07566.
\bibitem{add8}
M. S. Rudner and L. S. Levitov, Topological transition in a non-Hermitian quantum walk, Phys. Rev. Lett. \textbf{102}, 065703 (2009).
\bibitem{add9}
T. Kitagawa, M. S. Rudner, E. Berg, and E. Demler, Exploring topological phases with quantum walks, Phys. Rev. A \textbf{82}, 033429 (2010).
\bibitem{add7}
D. Bagrets, K. W. Kim, S. Barkhofen, S. De, J. Sperling, C. Silberhorn, A. Altland, and T. Micklitz, Probing the topological Anderson transition with quantum walks, Phys. Rev. Research \textbf{3}, 023183 (2021).



\bibitem{ref47}
S. Barkhofen, T. Nitsche, F. Elster, L. Lorz, A. Gábris, I. Jex, and C. Silberhorn, Measuring topological invariants in disordered discrete-time quantum walks, Phys. Rev. A \textbf{96}, 033846 (2017).
\bibitem{ref48}
X. Zhan, L. Xiao, Z. Bian, K. Wang, X. Qiu, B. C. Sanders, W. Yi, and P. Xue, Detecting topological invariants in nonunitary discrete-time quantum walks, Phys. Rev. Lett. \textbf{119}, 130501 (2017).
\bibitem{ref49}
T. Rakovszky, J. K. Asb\'{o}th, and A. Alberti, Detecting topological invariants in chiral symmetric insulators via losses, Phys. Rev. B \textbf{95}, 201407(R) (2017).
\bibitem{ref50}
F. Cardano, A. D'Errico, A. Dauphin, M. Maffei, B. Piccirillo, C. de Lisio, G. De Filippis, V. Cataudella, E. Santamato, L. Marrucci, M. Lewenstein, and P. Massignan, Detection of zak phases and topological invariants in a chiral quantum walk of twisted photons, Nature Comm. \textbf{8}, 15516 (2017).

\bibitem{add10}
D. Xie, T.-S. Deng, T. Xiao, W. Gou, T. Chen, W. Yi, and B. Yan, Topological quantum walks in momentum space with a Bose-Einstein condensate, Phys. Rev. Lett. \textbf{124}, 050502 (2020).


\bibitem{ref52}
A. Schreiber, K. N. Cassemiro, V. Poto\v{c}ek, A. G\'{a}bris, I. Jex, and Ch. Silberhorn, Decoherence and disorder in quantum walks: from ballistic spread to localization, Phys. Rev. Lett. \textbf{106}, 180403 (2011).
\bibitem{ref54}
H. Lavi\v{c}ka, V. Poto\v{c}ek, T. Kiss, E. Lutz, and I. Jex, Quantum walk with jumps, Eur. Phys. J. D \textbf{64}, 119 (2011).
\bibitem{ref51}
J. Svozil\'{i}k, R. de J. Le\'{o}n-Montiel, and J. P. Torres, Implementation of a spatial two-dimensional quantum random walk with tunable decoherence, Phys. Rev. A \textbf{86}, 052327 (2012).
\bibitem{ref53}
M. A. Pires and S. M. D. Queir\'{o}s, Quantum walks with sequential aperiodic jumps, Phys. Rev. E \textbf{102}, 012104 (2020).

\bibitem{ref58}
T. Oka, N. Konno, R. Arita, and H. Aoki, Breakdown of an electric-field driven system: a mapping to a quantum walk, Phys. Rev. Lett. \textbf{94}, 100602 (2005).
\bibitem{ref55}
R. J. Sension and R. J. Sension, Quantum path to photosynthesis, Nature (London) \textbf{446}, 740 (2007).
\bibitem{ref56}
M. Mohseni, P. Rebentrost, S. Lloyd, and A. Aspuru-Guzik, Environment-assisted quantum walks in photosynthetic energy Transfer, J. of Chem. Phys. \textbf{129}, 174106 (2008).


\bibitem{ref77}
A. Aharony, Spin-flop multicritical points in systems with random fields and in spin glasses, Phys. Rev. B \textbf{18}, 3328 (1978).
 \bibitem{ref76}
G. Misguich and C. Lhuillier, Two-dimensional quantum antiferromagnets, in frustrated spin systems, edited by H. T. Diep (World-Scientific, Singapore, 2005).
\bibitem{ref78}
 J. Villain, R. Bidaux, J.-P. Carton, and R. Conte, Order as an effect of disorder, J. Physique \textbf{41}, 1263 (1980);
 B. J. Minchau and R. A. Pelcovits, Two-dimensional $XY$ model in a random uniaxial field, Phys. Rev. B \textbf{32}, 3081 (1985); C. L. Henley, Ordering due to disorder in a frustrated vector antiferromagnet, Phys. Rev. Lett. \textbf{62}, 2056 (1989); 
 A. Moreo, E. Dagotto, T. Jolicoeur, and J. Riera, $Cu^{63}$ Knight shifts in the superconducting state of $YBa_2Cu_3$ $O_{7-\delta}$($T_c$=90 K), Phys. Rev. B \textbf{42}, 6283 (1990);
 D. E. Feldman, Exact zero-temperature critical behaviour of the ferromagnet in the uniaxial random field, J. Phys. A \textbf{31}, L177 (1998); G. E. Volovik, Random anisotropy disorder in superfluid 3He-A in aerogel, JETP Lett. \textbf{84}, 455 (2006);
 D. A. Abanin, P. A. Lee, and L. S. Levitov, Randomness-induced $XY$ ordering in a graphene quantum hall ferromagnet, Phys. Rev. Lett. \textbf{98}, 156801 (2007); L. Adamska, M. B. Silva Neto, and C. Morais Smith, $ La_{2-x}Sr_xCu_{1-z}Zn_{z}O_{4}$: Role of Dzyaloshinskii-Moriya and $XY$ anisotropies, Phys. Rev. B \textbf{75}, 134507 (2007);
\bibitem{ref80}
L. F. Santos, G. Rigolin, and C. O. Escobar, Entanglement versus chaos in disordered spin chains, Phys. Rev. A \textbf{69}, 042304 (2004); C. Mej\'{i}a-Monasterio, G. Benenti, G. G. Carlo, and G. Casati, Entanglement across a transition to quantum chaos, Phys. Rev. A \textbf{71}, 062324 (2005);
A. Lakshminarayan and V. Subrahmanyam, Multipartite entanglement in a one-dimensional time-dependent Ising model, Phys. Rev. A \textbf{71}, 062334 (2005); J. Karthik, A. Sharma, and A. Lakshminarayan, Entanglement, avoided crossings, and quantum chaos in an Ising model with a tilted magnetic field, Phys. Rev. A \textbf{75}, 022304 (2007);
W. G. Brown, L. F. Santos, D. J. Sterling, and L. Viola, Quantum chaos, delocalization, and entanglement in disordered heisenberg models, Phys. Rev. E \textbf{77}, 021106 (2008); 
F. Dukesz, M. Zilbergerts, and L. F. Santos, Interplay between interaction and (un)correlated disorder in one-dimensional many-particle systems: delocalization and global entanglement, New J. Phys. \textbf{11}, 043026 (2009); J. Hide, W. Son, and V. Vedral, Enhancing the detection of natural thermal entanglement with disorder, Phys. Rev. Lett. \textbf{102}, 100503 (2009); K. Fujii and K. Yamamoto, Anti-Zeno effect for quantum transport in disordered systems, Phys. Rev. A \textbf{82}, 042109 (2010).

\bibitem{ref81}
R. Prabhu, S. Pradhan, A. Sen(De), and U. Sen, Disorder overtakes order in information concentration over quantum networks, Phys. Rev. A \textbf{84}, 042334 (2011).
\bibitem{ref79}
P. V. Mart\'{i}n, J. A. Bonachela, and M. A. Mu\~{n}oz, Quenched disorder forbids discontinuous transitions in nonequilibrium low-dimensional systems, Phys. Rev. E \textbf{89}, 012145 (2014).
\bibitem{ref82}
A. Niederberger, T. Schulte, J. Wehr, M. Lewenstein, L. Sanchez-Palencia, and K. Sacha, Disorder-induced order in two-component Bose-Einstein condensates, Phys. Rev. Lett. \textbf{100}, 030403 (2008); A. Niederberger, J. Wehr, M. Lewenstein, and K. Sacha, Disorder-induced phase control in superfluid Fermi-Bose mixtures, Europhys. Letts. \textbf{86}, 26004 (2009); A. Niederberger, M. M. Rams, J. Dziarmaga, F. M. Cucchietti, J. Wehr, and M. Lewenstein, Disorder-induced order in quantum $XY$ chains, Phys. Rev. A \textbf{82}, 013630 (2010); D. I. Tsomokos, T. J. Osborne, and C. Castelnovo, Interplay of topological order and spin glassiness in the toric code under random magnetic fields, Phys. Rev. B \textbf{83}, 075124 (2011); M. S. Foster, H.-Y. Xie, and Y.-Z. Chou, Topological protection, disorder, and interactions: survival at the surface of three-dimensional topological superconductors, Phys. Rev. B \textbf{89}, 155140 (2014).



\bibitem{ref104}
P. Ribeiro, P. Milman, and R. Mosseri, Aperiodic quantum random walks, Phys. Rev. Lett. \textbf{93}, 190503 (2004).
\bibitem{ref105}
M. C. Ba\~{n}uls, C. Navarrete, A. P\'{e}rez, E. Rold\'{a}n, and J. C. Soriano, Quantum walk with a time-dependent coin, Phys. Rev. A \textbf{73}, 062304 (2006).
\bibitem{ref100}
D. Bulger, J. Freckleton, and J. Twamley, Position-dependent and cooperative quantum Parrondo walks, New J. Phys. \textbf{10}, 093014 (2008).
\bibitem{ref106}
A. Romanelli, The fibonacci quantum walk and its classical trace map, Physica A \textbf{388}, 3985 (2009).
\bibitem{ref107}
 A. Romanelli, Driving quantum-walk spreading with the coin operator, Phys. Rev. A \textbf{80}, 042332 (2009).
\bibitem{ref101}
Y. Shikano and H. Katsura, Localization and fractality in inhomogeneous quantum walks with self-duality, Phys. Rev. E \textbf{82}, 031122 (2010).
\bibitem{ref62}
 C. M. Chandrashekar, Disordered-quantum-walk-induced localization of a Bose-Einstein condensate, Phys. Rev. A \textbf{83}, 022320 (2011).
 \bibitem{ref89}
C. Ampadu, Limit Theorems for the Disordered quantum Walk, arXiv:1108.6110.
\bibitem{ref90}
A. Ahlbrecht, V. B. Scholz, and A. H. Werner, disordered quantum walks in one lattice dimension, J. Math. Phys. \textbf{52}, 102201 (2011).
\bibitem{ref93}
C. M. Chandrashekar, Disordered quantum walk-induced localization of a Bose-Einstein condensate, Phys. Rev. A \textbf{83}, 022320 (2011).
\bibitem{ref94}
A. Ahlbrecht, C. Cedzich, R. Matjeschk, V. B. Scholz, A. H. Werner, and R. F. Werner, Asymptotic behavior of quantum walks with spatio-temporal coin fluctuations, Quantum Inf. Processing \textbf{11}, 1219 (2012). 
 \bibitem{ref83}
C. M. Chandrashekar, Disorder induced localization and enhancement of entanglement in one- and two-dimensional quantum walks, arXiv:1212.5984.
 \bibitem{ref61}
R. Vieira, E. P. M. Amorim, and G. Rigolin, Dynamically disordered quantum walk as a maximal entanglement generator, Phys. Rev. Lett. \textbf{111}, 180503 (2013).
\bibitem{ref102}
N. Konno, T. {\L}uczak, and E. Segawa, Limit measures of inhomogeneous discrete-time quantum walks in one dimension, Quantum Inf. Processing  \textbf{12}, 33 (2013).
\bibitem{ref103}
R. Zhang, P. Xue, and J. Twamley, One-dimensional quantum walks with single-point phase defects,  Phys. Rev. A \textbf{89}, 042317
(2014).
\bibitem{ref60}
R. Vieira, E. P. M. Amorim, and G. Rigolin, Entangling power of disordered quantum walks, Phys. Rev. A  \textbf{89}, 042307 (2014). 
\bibitem{ref46}
B. Tarasinski, J. K. Asbóth, and J. P. Dahlhaus, Scattering theory of topological phases in discrete-time quantum walks, Phys. Rev. A \textbf{89}, 042327 (2014).
\bibitem{ref91}
C. M. Chandrashekar and T. Busch, Quantum percolation and transition point of a directed discrete-time quantum walk, Sci. Rep. \textbf{4}, 6583 (2014).
 \bibitem{ref108}
 M. Montero, Invariance in quantum walks with time-dependent coin operators, Phys. Rev. A \textbf{90}, 062312 (2014).
 \bibitem{ref65}
 Q. Zhao  and J. Gong, From disordered quantum walk to physics of off-diagonal disorder, Phys. Rev. B \textbf{92}, 214205 (2015).
 \bibitem{ref66}
 T. Rakovszky and J. K. Asboth, Localization, delocalization, and topological phase transitions in the one-dimensional split-step quantum walk, Phys. Rev. A \textbf{92}, 052311 (2015).
 \bibitem{ref96}
M. Montero, Classical-like behavior in quantum walks with inhomogeneous, time-dependent coin operators, Phys. Rev. A \textbf{93}, 062316 (2016).
 \bibitem{ref87}
S. Chakraborty, A. Das, A. Mallick, and C. M. Chandrashekar, Quantum ratchet in disordered quantum walk, Annalen der Physik, \textbf{529}, 8, 1600346 (2017).
\bibitem{ref88}
K. Mochizuki, H. Obuse, Effects of disorder on non-unitary $\mathcal{PT}$ symmetric quantum walks, Interdisciplinary Information Sciences \textbf{23}, 95 (2017).
\bibitem{ref92}
A. D. Verga, Edge states in a two-dimensional quantum walk with disorder, Eur. Phys. J. B \textbf{90}, 41 (2017).
\bibitem{ref84}
S. Singh and C. M. Chandrashekar, Quantum interference and coherence in one-dimensional disordered and localized quantum walk, arXiv:1711.06217.
 \bibitem{ref64}
 N. P. Kumar, S. Banerjee, and C. M. Chandrashekar, Enhanced non-markovian behavior in quantum walks with markovian disorder, Sci. Rep. \textbf{8}, 8801 (2018).
 \bibitem{ref97}
G. D. Molfetta, D. O. Soares-Pinto, and S. M. D. Queir\'{o}s, Elephant quantum walk, Phys. Rev. A \textbf{97}, 062112 (2018).
\bibitem{ref59}
A. C. Orthey Jr. and E. P. M. Amorim, Weak disorder enhancing the production of entanglement in quantum walks, Braz. J. Phys. \textbf{49}, 595 (2019).
\bibitem{ref63}
 S. Das, S. Mal, A. Sen, and U. Sen, Inhibition of spreading in quantum random walks due to quenched Poisson-distributed disorder, Phys. Rev. A \textbf{99}, 042329 (2019).
\bibitem{ref86}
M. A. Pires and S. M. D. Queir\'{o}s, Negative correlations can play a positive role in disordered quantum walks, arXiv:2008.08867.
\bibitem{ref95}
M. A. Pires and S. M. D. Queir\'{o}s, Genuine Parrondo's paradox in quantum walks with time-dependent coin operators, Phys. Rev. E \textbf{102}, 042124 (2020).
\bibitem{ref99}
P. Sen, Scaling and crossover behaviour in a truncated long range quantum walk, Physica A \textbf{545}, 123529 (2020).
\bibitem{ref85}
F. Nosrati, A. Laneve, M. K. Shadfar, A. Geraldi, K. Mahdavipour, F. Pegoraro, P. Mataloni, and R. L. Franco, Quantum information spreading in a disordered quantum walk, J. Opt. Soc. Am. B \textbf{38}, 2570 (2021).
\bibitem{ref98}
D. Bagrets, K. W. Kim, S. Barkhofen, S. De, J. Sperling, C. Silberhorn, A. Altland, and T. Micklitz, Probing the topological anderson transition with quantum walks, Phys. Rev. Research \textbf{3}, 023183 (2021).






 
 
 
 \bibitem{ref109}
M. Kac, Random  walk  and  the  theory  of  Brownian  motion, The American Mathematical Monthly, \textbf{54}, 369 (1947). 
\bibitem{ref110}
F. B. Knight, On the random walk and Brownian motion, Trans. Amer. Math. Soc. \textbf{103}, 218 (1962).
\bibitem{ref111}
J. {\AA}berg, Quantifying superposition, arXiv:quant-ph/0612146.
\bibitem{ref112}
T. Baumgratz, M. Cramer, and M. B. Plenio, Quantifying
coherence, Phys. Rev. Lett. \textbf{113}, 140401 (2014).
\bibitem{ref113}
A. Winter and D. Yang, Operational resource theory of
coherence, Phys. Rev. Lett. \textbf{116}, 120404 (2016).
\bibitem{ref114}
A. Streltsov, G. Adesso, and M. B. Plenio, Quantum
coherence as a resource, Rev. Mod. Phys. \textbf{89}, 041003
(2017).
\bibitem{ref115}
T. Theurer, N. Killoran, D. Egloff, and M. B. Plenio, Resource theory of superposition, Phys. Rev. Lett. \textbf{119}, 230401 (2017); S. Das, C. Mukhopadhyay, S. S. Roy, S. Bhattacharya, A. Sen(De), and U. Sen, Wave-particle duality employing quantum coherence in superposition with non-orthogonal pointers, J. Phys. A: Math. Theor. \textbf{53}, 115301 (2020).
\bibitem{ref116}
 C. Srivastava, S. Das, and U. Sen, Resource theory of quantum coherence with probabilistically nondistinguishable pointers and corresponding wave-particle duality, Phys. Rev. A \textbf{103}, 022417 (2021).
\bibitem{ref117}
I. Banerjee, K. Sen, C. Srivastava, and U. Sen, Quantum
coherence with incomplete set of pointers and corresponding wave-particle duality, arXiv:2108.05849.
\bibitem{ref118}
M. B. Plenio and S. Virmani, An introduction to entanglement measures, Quant. Inf. Comput. \textbf{7}, 1 (2007).
\bibitem{ref119}
R. Horodecki, P. Horodecki, M. Horodecki, and K.
Horodecki, Quantum entanglement, Rev. Mod. Phys. \textbf{81},
865 (2009).
\bibitem{ref120}
O. G\"{u}hne and G. T\'{o}th, Entanglement detection, Physics
Reports \textbf{474}, 1 (2009).
\bibitem{ref121}
S. Das, T. Chanda, M. Lewenstein, A. Sanpera, A.
Sen(De), and U. Sen, The separability versus entanglement problem, in Quantum Information, edited by D. Bru{\ss} and G. Leuchs (Wiley, Weinheim, 2019), chapter 8.
\bibitem{ref73}
R. M. Clark and B. J. Morrison, A normal approximation to the Fisher distribution, Geophys. J. R. Astr. Soc. \textbf{13}, 271 (1983).
\bibitem{ref74}
T. D. Downs, Cauchy families of directional distributions closed under location and scale transformations, 
The Open Statistics \& Probability Journal, \textbf{1}, 76 (2009). 
\bibitem{ref75}
S. Kato and P. McCullagh, M\"{o}bius transformation and a Cauchy family on the sphere, Bernoulli, \textbf{26}, 3224 (2020).

\end{thebibliography}
\end{document}